\documentclass[aps,pra,reprint,amsmath,amssymb,showpacs,floatfix]{revtex4-1}

\usepackage[colorlinks,allcolors=blue,hyperindex,breaklinks]{hyperref}

\usepackage{graphicx,units}

\newcommand{\ket}[1]{\ensuremath{\vert{#1\rangle}}} 
\newcommand{\bra}[1]{\ensuremath{{\langle #1}\vert}}

\newcommand{\op}[1]{\hat{#1}}

\newcommand{\I}{\text{i}}
\newcommand{\E}{\text{e}}

\newcommand{\buvec}[1]{\ensuremath{\mathbf{\hat{#1}}}}

\newcommand{\bopvecgr}[1]{\ensuremath{\mathbf{\op{\boldsymbol #1}}}}

\begin{document}

\title{Protective measurement of a qubit by a qubit probe}

\author{Maximilian Schlosshauer}

\affiliation{Department of Physics, University of Portland, 5000 North Willamette Boulevard, Portland, Oregon 97203, USA}

\begin{abstract}
We study the protective measurement of a qubit by a second qubit acting as a probe. Consideration of this model is motivated by the possibility of its experimental implementation in multiqubit systems such as trapped ions. In our scheme, information about the expectation value of an arbitrary observable of the system qubit is encoded in the rotation of the state of the probe qubit. We describe the structure of the Hamiltonian that gives rise to this measurement and analyze the resulting dynamics under a variety of realistic conditions, such as noninfinitesimal measurement strengths, repeated measurements, non-negligible intrinsic dynamics of the probe, and interactions of the system and probe qubits with an environment. We propose an experimental realization of our model in an ion trap. The experiment may be performed with existing technology and makes use of established experimental methods for the engineering and control of Hamiltonians for quantum gates and quantum simulations of spin systems. \\[.2cm]
Journal reference: \emph{Phys.\ Rev.\ A} {\bf 101}, 042113 (2020),  DOI: \href{https://doi.org/10.1103/PhysRevA.101.042113}{\texttt{10.1103/PhysRevA.101.042113}}
\end{abstract}

\maketitle

\section{Introduction}

Weak quantum measurements have attracted widespread theoretical and experimental interest \cite{Dressel:2014:uu,Gao:2014:cu}. In contrast with standard impulsive measurements, they allow one to obtain information about a quantum system without appreciably affecting its state during the measurement. An important instance of such weak measurements is protective measurement \cite{Aharonov:1993:qa,Aharonov:1993:jm,Dass:1999:az,Vaidman:2009:po,Gao:2014:cu,Genovese:2017:zz,Qureshi:2015:jj}. Here, the state of the system is prevented from changing during the measurement by preparing the system in an eigenstate of a self-Hamiltonian that is much stronger than the weak-interaction Hamiltonian describing the coupling of the system to the measurement probe. The interaction between system and probe is for a duration $T$ much larger than the timescale set by the intrinsic evolution of the system. If these conditions are fulfilled, one can show that the pointer of the probe is shifted such as to indicate the expectation value of an arbitrary observable of the system (the particular observable is determined by the structure of the interaction Hamiltonian). Thus, the expectation value, a quantity usually obtained statistically from measurements on an ensemble of systems, can be obtained in a single-shot measurement on an individual system without appreciably disturbing the state of the system \cite{Aharonov:1993:qa,Aharonov:1993:jm,Dass:1999:az,Vaidman:2009:po,Gao:2014:cu,Genovese:2017:zz,Qureshi:2015:jj}. This suggests the possibility of quantum-state measurement of single systems \cite{Aharonov:1993:qa,Aharonov:1993:jm,Aharonov:1996:fp,Dass:1999:az,Vaidman:2009:po,Auletta:2014:yy,Diosi:2014:yy,Aharonov:2014:yy,Schlosshauer:2016:uu}, as well as a number of other applications in quantum measurement \cite{Aharonov:1993:qa,Aharonov:1993:jm,Aharonov:1996:fp,Alter:1997:oo,Dass:1999:az,Gao:2014:cu} and the study of particle trajectories \cite{Aharonov:1996:ii,Aharonov:1999:uu}. 

Despite the recognition of the importance of protective measurements, their experimental realization using the scheme just described has remained an open challenge. The paradigmatic example frequently considered in models of protective measurements is that of a setup of the Stern--Gerlach type, in which a spin-$\frac{1}{2}$ particle is deflected by an inhomogeneous magnetic field while an additional, much stronger uniform field provides the protection of the spin state  \cite{Aharonov:1993:jm,Dass:1999:az,Schlosshauer:2015:uu}. For parameters typical to Stern--Gerlach experiments, however, achieving both sufficient state protection and appreciable beam displacements requires a very strong uniform field of several Tesla \cite{Schlosshauer:2015:uu} (unless very slow, cold atoms are used \cite{Dass:1999:le}), and the fields need to be extended over a sizable region of space (on the order of \unit[0.1--1]{m}), posing an experimentally highly challenging scenario. If one were instead to use photons to implement the protective measurement, the difficulty lies in applying both the protection and measurement Hamiltonians simultaneously, as individual optical elements such as birefringent plates can only realize one of these Hamiltonians. (There exists a different version of a protective measurement based on the quantum Zeno effect \cite{Aharonov:1993:jm} that has been realized using photons \cite{Piacentini:2017:oo}. However, because the state protection is realized through repeated projections onto the initial state, it requires \emph{a priori} knowledge of this state and thus precludes measurement of an unknown quantum state for a single system \footnote{In the standard scheme used in this paper, Alice can give a protected state (that she may know) to Bob, who can then apply the measurement interaction to obtain expectation values without needing to know the protection procedure or the protected state. In this sense, Bob can measure an unknown quantum state. By contrast, in the quantum Zeno version, Bob needs to apply projections on the state throughout the measurement interaction, and would thus need to know the state all along.}. In the following, we will take the term ``protective measurement'' to refer to the non-Zeno scheme described in the preceding paragraph.)

The current impasse in the experimental realization of a protective measurement suggests the search for alternative implementations. Here we propose and analyze protective qubit measurements in which the probe is realized by a two-level system implemented by a second qubit (to be referred to as a ``qubit probe'' from here on). This is in contrast with existing treatments of protective measurements \cite{Aharonov:1993:qa,Aharonov:1993:jm,Dass:1999:az,Vaidman:2009:po,Gao:2014:cu,Genovese:2017:zz,Qureshi:2015:jj,Schlosshauer:2014:pm,Schlosshauer:2015:uu}, where the pointer shift is encoded in a translation in position or momentum of a particle moving in phase space (henceforth referred to as a ``phase-space probe''). One benefit of a qubit probe is that one can make use of the many experimentally well-established techniques for engineered interactions between qubits. Specifically, experiments with trapped ions \cite{Bruzewicz:2019:aa}, both those aimed at quantum computation \cite{Bruzewicz:2019:aa} and those designed to simulate many-spin systems \cite{Milburn:2000:az,Porras:2004:tt,Lin:2011:aa,Korenblit:2012:zz,Jurcevic:2014:uu,Hayes:2014:kk,Smith2016:im,Monroe:2019:za}, are able to realize a wide variety of single- and multiqubit Hamiltonians, which can be directly applied to the implementation of a protective measurement. Moreover, such experiments also provide fast, high-fidelity state preparation and readout \cite{Noek:2013:uu,Harty:2014:rr,Bruzewicz:2019:aa}. Thus our model offers the possibility of an experimental implementation of a tunable protective measurement with existing technology. 

This paper is organized as follows. In Sec.~\ref{sec:model}, we describe the model and its Hamiltonian, solve for the resulting dynamics, and discuss the readout of the qubit probe. In Sec.~\ref{sec:nonid-meas}, we study our model under realistic conditions, such as noninfinitesimal measurement strengths, repeated measurements, intrinsic probe dynamics, and interactions with an environment during the measurement. In Sec.~\ref{sec:prop-exper-impl}, we propose an experimental implementation of the model with trapped ions. We discuss our findings in Sec.~\ref{sec:discussion}.

\section{\label{sec:model}Model and dynamics}

\subsection{\label{sec:hamiltonian}Hamiltonian and time evolution}

The Hamiltonian describing the protective measurement of a system $S$ by a probe $P$ takes the general form \cite{Aharonov:1993:qa,Aharonov:1993:jm,Dass:1999:az,Vaidman:2009:po,Gao:2014:cu}
\begin{equation}\label{eq:3aaa}
\op{H}(t) = \op{H}_S+\op{H}_P+\op{H}_m(t) = \op{H}_S+ \op{H}_P + \kappa(t) \op{O}_S \otimes \op{O}_P,
\end{equation}
where $\op{H}_S$ and $\op{H}_P$ are the self-Hamiltonians of $S$ and $P$, and $\op{H}_m(t)$ represents the measurement interaction. $\op{O}_S$ is an arbitrary observable of $S$ that is to be measured,  $\op{O}_P$ is an operator that generates the shift of the probe pointer, and $\kappa(t)$ represents the time dependence of the measurement interaction. Commonly, one takes $\kappa(t) \propto 1/T$ during the measurement interval $t \in [0,T]$ and $\kappa(t)=0$ otherwise \cite{Aharonov:1993:qa,Aharonov:1993:jm,Dass:1999:az,Vaidman:2009:po,Gao:2014:cu,Schlosshauer:2014:pm,Schlosshauer:2015:uu}. Then the Hamiltonian is time-independent throughout the duration of the measurement. One assumes that $S$ starts in an eigenstate of $\op{H}_S$ at $t=0$, with the result of the measurement  evaluated at $t=T$. It is customary to neglect the self-Hamiltonian $\op{H}_P$ of $P$, such that the evolution of $P$ is entirely due to the coupling to $S$ (we will relax this assumption in Sec.~\ref{sec:infl-intr-probe} below). 

We focus on the case of a qubit system, with $\op{H}_S=\frac{1}{2}\hbar \omega_0 \op{\sigma}_z$ and arbitrary qubit observable $\op{O}_S = \bopvecgr{\sigma} \cdot \buvec{m}$, where we express the unit vector $\buvec{m}$ in terms of polar and azimuthal angles $\gamma$ and $\eta$, $\buvec{m} = (\cos\eta\sin\gamma, \sin\eta\sin\gamma, \cos\gamma)$. We consider a probe represented by a qubit acting as an ancilla (for a study of the use of a qubit probe in weak-value measurements \cite{Dressel:2014:uu}, see Ref.~\cite{Wu:2009:zz}). Then the pointer observable is $\op{O}_P = \bopvecgr{\sigma} \cdot \buvec{n}$, which generates rotations of the state of $P$ on the Bloch sphere around the $\buvec{n}$ axis. Thus, the pointer shift is represented by a qubit rotation, and the Hamiltonian governing the evolution of the system and probe qubits during the measurement interval $t \in [0,T]$ is
\begin{equation}\label{eq:vdkkjl7}
\op{H} =\op{H}_S+\op{H}_m = \frac{1}{2}\hbar \omega_0 \op{\sigma}_z + \frac{\hbar\lambda}{T} (\bopvecgr{\sigma} \cdot \buvec{m}) \otimes (\bopvecgr{\sigma} \cdot \buvec{n}),
\end{equation}
where $\lambda$ is a dimensionless constant of order unity that we will fix below. 

The evolution given by this Hamiltonian can be solved exactly. For each of the two eigenstates $\ket{\pm}_\buvec{n}$ (with eigenvalues $\pm 1$) of the probe part $\bopvecgr{\sigma} \cdot \buvec{n}$ of the Hamiltonian, we can consider a corresponding effective Hamiltonian $\op{H}_\pm$ for the system qubit $S$ given by
\begin{equation}\label{eq:vdvdvqqdkkjl7}
\op{H}_\pm = \frac{1}{2}\hbar \omega_0 \op{\sigma}_z \pm \frac{\hbar\lambda}{T} (\bopvecgr{\sigma} \cdot \buvec{m}) \equiv \frac{1}{2}\hbar \omega_0(\bopvecgr{\sigma} \cdot \buvec{w}^\pm).
\end{equation}
The components of $\buvec{w}^\pm$ are 
\begin{subequations}\label{eq:26dvrr}
\begin{align}
w_x^\pm &= \pm 2\lambda\xi\cos\eta\sin\gamma, \\
w_y^\pm &= \pm 2\lambda\xi \sin\eta\sin\gamma, \\
w_z^\pm &= 1 \pm 2\lambda \xi \cos\gamma,
\end{align}
\end{subequations}
where the dimensionless constant $\xi = (\omega_0 T)^{-1}$ measures the strength of the measurement interaction $\op{H}_m$ relative to the protection Hamiltonian $\op{H}_S$. The magnitude of $\buvec{w}^\pm$ is 
\begin{align}\label{eq:ko4}
\chi_\pm& = \bigl[ 1 + (2\lambda\xi)^2  \pm 4 \lambda\xi \cos\gamma \bigr]^{1/2}.
\end{align}
The eigenstates of the Hamiltonian $\op{H}_\pm$ defined in Eq.~\eqref{eq:vdvdvqqdkkjl7} are 
\begin{subequations}
\begin{align}
\ket{\phi_0^\pm} &= \cos\frac{\theta_\pm}{2}\ket{0} + \sin\frac{\theta_\pm}{2}\E^{\I \phi_\pm}\ket{1},\\
\ket{\phi_1^\pm} &= \sin\frac{\theta_\pm}{2}\ket{0} - \cos\frac{\theta_\pm}{2}\E^{\I \phi_\pm}\ket{1},
\end{align}
\end{subequations}
where $\theta_\pm$ and $\phi_\pm$ are the polar and azimuthal angles of $\buvec{w}^\pm$ (where $\cos\theta_\pm = w_z^\pm \chi_\pm^{-1}$), and the eigenvalues are $E_0^\pm = + \frac{1}{2}\hbar \omega_0\chi_\pm$ and $E_1^\pm = - \frac{1}{2}\hbar \omega_0\chi_\pm$. 

Following the protective-measurement protocol, we take the initial state of $S$ (at $t=0$) to be the eigenstate $\ket{0}$ of $\op{H}_S$, where we may write $\ket{0}=\cos\frac{\theta_\pm}{2}\ket{\phi_0^\pm}+ \sin\frac{\theta_\pm}{2}\ket{\phi_1^\pm}$. Then, for an arbitrary pure initial probe state $\ket{\psi_P(0)}=c_+ \ket{+}_\buvec{n} + c_- \ket{-}_\buvec{n}$, the final system--probe state at $t=T$ is 
\begin{align}\label{eq:vihdgaass7cf6gv}
\ket{\Psi(T)} &= 
c_+ \biggl[ \E^{ - \I \omega_0\chi_+ T/2}  \cos\frac{\theta_+}{2}\ket{\phi_0^+} \notag\\ & \quad + \E^{ + \I \omega_0\chi_+ T/2} \sin\frac{\theta_+}{2}\ket{\phi_1^+} \biggr]\ket{+}_\buvec{n} \notag\\ &\quad
+ c_- \biggl[ \E^{ - \I \omega_0\chi_- T/2}  \cos\frac{\theta_-}{2}\ket{\phi_0^-}\notag\\ & \quad + \E^{ + \I \omega_0\chi_- T/2} \sin\frac{\theta_-}{2}\ket{\phi_1^-} \biggr]\ket{-}_\buvec{n}.
\end{align}
In an ideal protective measurement, the measurement interaction is weak compared to $\op{H}_S$, i.e., $\xi \ll 1$. Then $\chi_\pm \approx 1\pm 2\lambda\xi \cos\gamma$ [see Eq.~\eqref{eq:ko4}] and $\theta_+ \approx \theta_- \ll 1$, and the state~\eqref{eq:vihdgaass7cf6gv} becomes
\begin{align}\label{eq:vihd554ass7cf6gv}
\ket{\Psi(T)} & \approx
\E^{ - \I \omega_0 T/2}\ket{0}  \left[c_+ \E^{ - \I \lambda\cos\gamma}  \ket{+}_\buvec{n} + c_- \E^{ +\I \lambda\cos\gamma} \ket{-}_\buvec{n} \right]\notag\\
&= \E^{ - \I \omega_0 T/2}\ket{0} \E^{ - \I \lambda \cos\gamma (\bopvecgr{\sigma} \cdot \buvec{n})}  \ket{\psi_P(0)}.
\end{align}
Therefore, the measurement interaction rotates the probe state around the $\buvec{n}$ axis on the Bloch sphere by an angle 
\begin{equation}\label{eq:7buhdjk56745687bj}
\Theta(\gamma) = 2\lambda \cos\gamma = 2\lambda  \bra{0} \bopvecgr{\sigma} \cdot \buvec{m}  \ket{0},
\end{equation}
where $\bra{0} \bopvecgr{\sigma} \cdot \buvec{m}  \ket{0}$ is the expectation value of $\bopvecgr{\sigma} \cdot \buvec{m}$ in the initial state of $S$. Thus, as expected from the general theory of ideal protective measurements \cite{Aharonov:1993:qa,Aharonov:1993:jm,Dass:1999:az,Vaidman:2009:po,Gao:2014:cu}, the probe pointer is shifted by an amount proportional to $\bra{0} \bopvecgr{\sigma} \cdot \buvec{m}  \ket{0}$ while the state of $S$ remains approximately unchanged. 

\subsection{\label{sec:probe-readout}Probe readout}

Our analysis suggests the following strategy for generating and measuring the pointer shift, i.e., the probe rotation (see also Ref.~\cite{Wu:2009:zz} for a similar scheme but applied to weak-value measurements \cite{Dressel:2014:uu}). We choose an arbitrary probe rotation axis $\buvec{n}$ and initialize the probe in an eigenstate of $\bopvecgr{\sigma} \cdot \buvec{n}_\perp$, where $\buvec{n}_\perp$ is a unit vector perpendicular to $\buvec{n}$. The interaction with the qubit system $S$ will then rotate the probe state out of the plane spanned by $\buvec{n}$ and $\buvec{n}_\perp$, such that the state acquires a component in the direction given by $\buvec{n} \times \buvec{n}_\perp$. We choose the constant $\lambda$ such that for the maximum value $\cos\gamma=1$ (which corresponds to $\bra{0} \op{\sigma}_z\ket{0}$), the rotation angle is such that the probe ends up in a state perpendicular to the plane spanned by $\buvec{n}$ and $\buvec{n}_\perp$, i.e., that it points in the direction given by $\buvec{k} = \buvec{n} \times \buvec{n}_\perp$. This implies a maximum rotation angle of $\pm \pi/2$ on the Bloch sphere and hence the choice $\lambda=\pi/4$.

Since the rotation angle $\Theta(\gamma)$ encodes the desired expectation value $\bra{0} \bopvecgr{\sigma} \cdot \buvec{m}  \ket{0}$ [see Eq.~\eqref{eq:7buhdjk56745687bj}], readout of the pointer corresponds to measuring this rotation angle. This can be done by measuring the expectation value of the observable $\bopvecgr{\sigma} \cdot \buvec{k}$ on the probe, which gives the component of the Bloch vector along the $\buvec{k}$ axis. For an ideal protective measurement, the total rotation angle around the $\buvec{n}$ axis is $\frac{\pi}{2}  \bra{0} \bopvecgr{\sigma} \cdot \buvec{m}  \ket{0}$ [see Eq.~\eqref{eq:7buhdjk56745687bj}], and thus in this case the corresponding pointer expectation value at $t=T$ is given by
\begin{equation}\label{eq:1}
\langle\bopvecgr{\sigma} \cdot \buvec{k}\rangle = \bra{\Psi(T)} (\op{I} \otimes \bopvecgr{\sigma} \cdot \buvec{k}) \ket{\Psi(T)} = \sin\left(\frac{\pi}{2}\bra{0} \bopvecgr{\sigma} \cdot \buvec{m}  \ket{0}\right),
\end{equation}
with $\ket{\Psi(T)}$ given by Eq.~\eqref{eq:vihd554ass7cf6gv}. We will refer to the expectation value~\eqref{eq:1} as the \emph{ideal value} from here on.

Measurement of $\langle\bopvecgr{\sigma} \cdot \buvec{k}\rangle$ may be accomplished by repeating the following procedure: (i) initialization of the probe $P$ in the state $\ket{\psi_P}$; (ii) interaction of $P$ with the system $S$ for a time $T$; and (iii) measurement of $\bopvecgr{\sigma} \cdot \buvec{k}$ on $P$ (see Sec.~\ref{sec:discussion} for a discussion and comparison of this readout to the case of a phase-space probe). Note that nonetheless only a single system $S$ is required, and its quantum state will remain approximately unchanged throughout the process. Thus, the essential feature of a protective measurement is preserved, namely, that it allows us to measure an expectation value for an individual system, with in principle arbitrarily small state disturbance (see also Secs.~\ref{sec:repeated} and \ref{sec:discussion} for further analysis).

Since we are free to select an arbitrary orientation of the rotation axis for the probe, in what follows we will choose, for ease of notation and visualization, the probe rotation to be about the $y$ axis. Thus the system--probe Hamiltonian~\eqref{eq:vdkkjl7} (with $\lambda=\pi/4$) assumes the form
\begin{equation}\label{eq:7vfdklkln}
\op{H} =\op{H}_S+\op{H}_m =\frac{1}{2}\hbar \omega_0 \op{\sigma}_z + \frac{\hbar}{T}\frac{\pi}{4} (\bopvecgr{\sigma} \cdot \buvec{m}) \otimes \op{\sigma}_y.
\end{equation}
We shall also take the initial probe state to be the eigenstate $\ket{0}$ of $\op{\sigma}_z$. Thus, the system and probe states are initially aligned. (In practice, one does not necessarily know the state of the system \cite{Aharonov:1993:jm,Dass:1999:az}, and so one would simply choose the initial probe state along some arbitrary axis perpendicular to the rotation axis, as discussed above.) Then the rotation angle of the probe is obtained by measuring the expectation value $\langle\op{\sigma}_x\rangle$ on the probe.

\begin{figure*}[ht!]
(a) \hspace{3.7cm} (b) \hspace{3.7cm} (c) 

\includegraphics[width=1.65in]{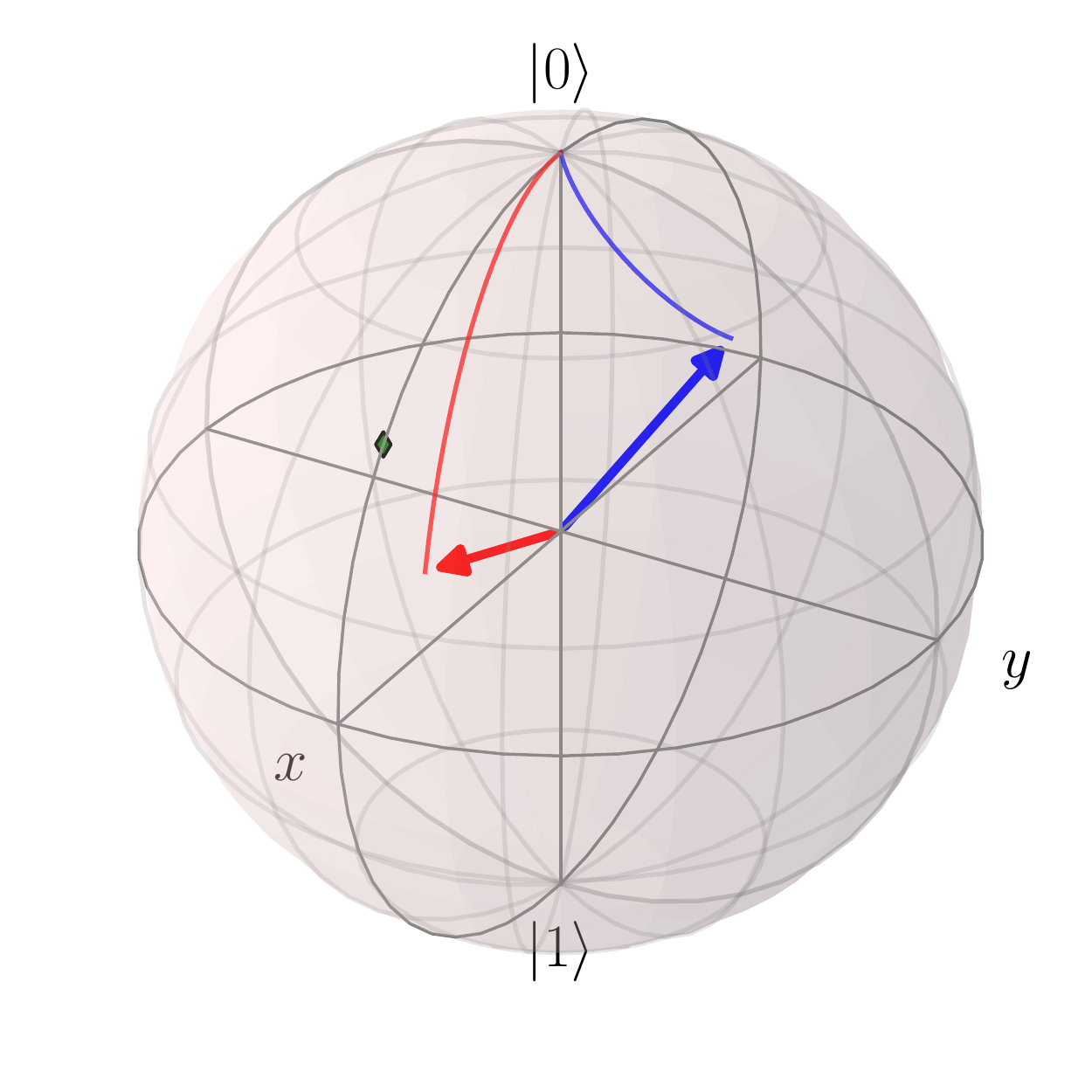} \includegraphics[width=1.65in]{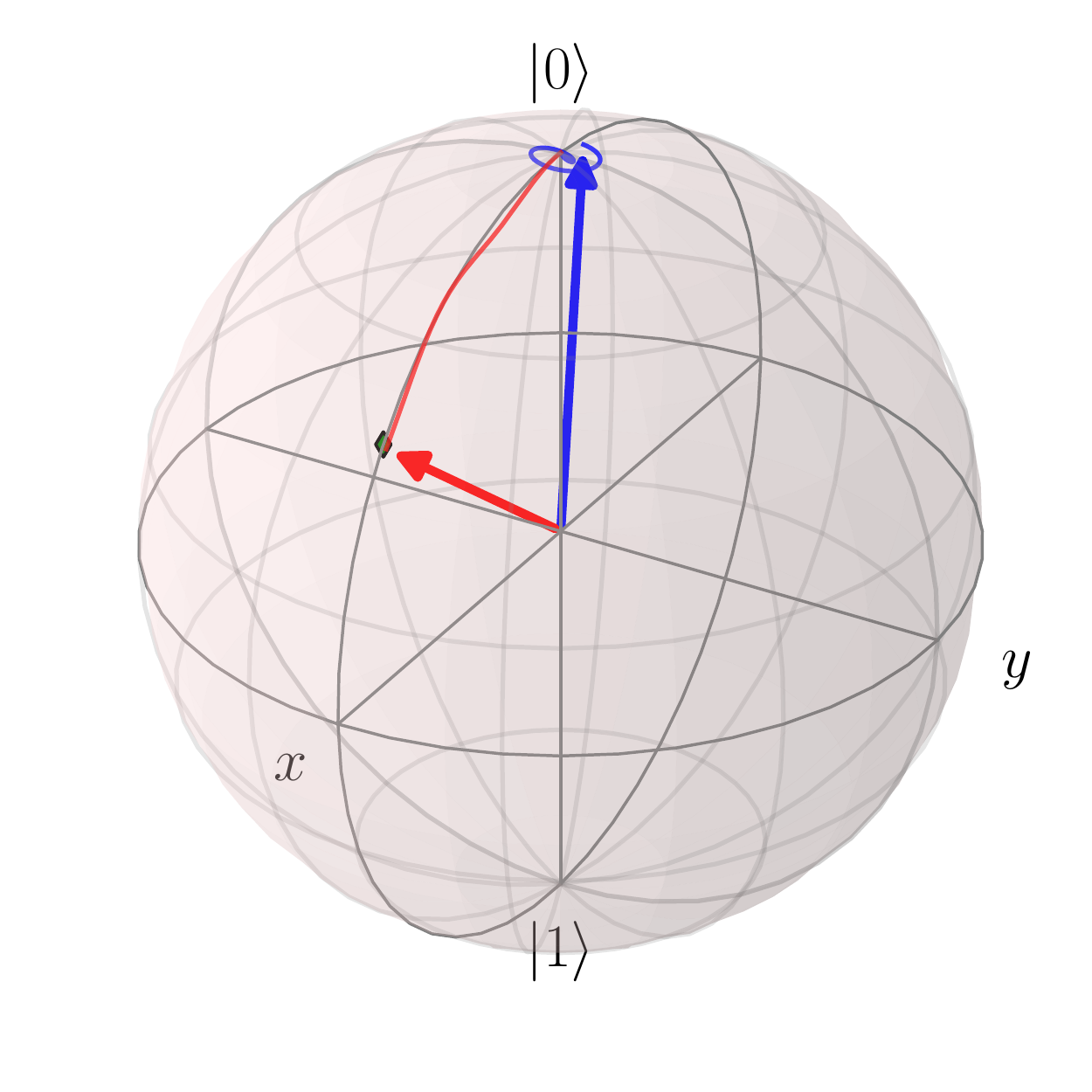} \includegraphics[width=1.65in]{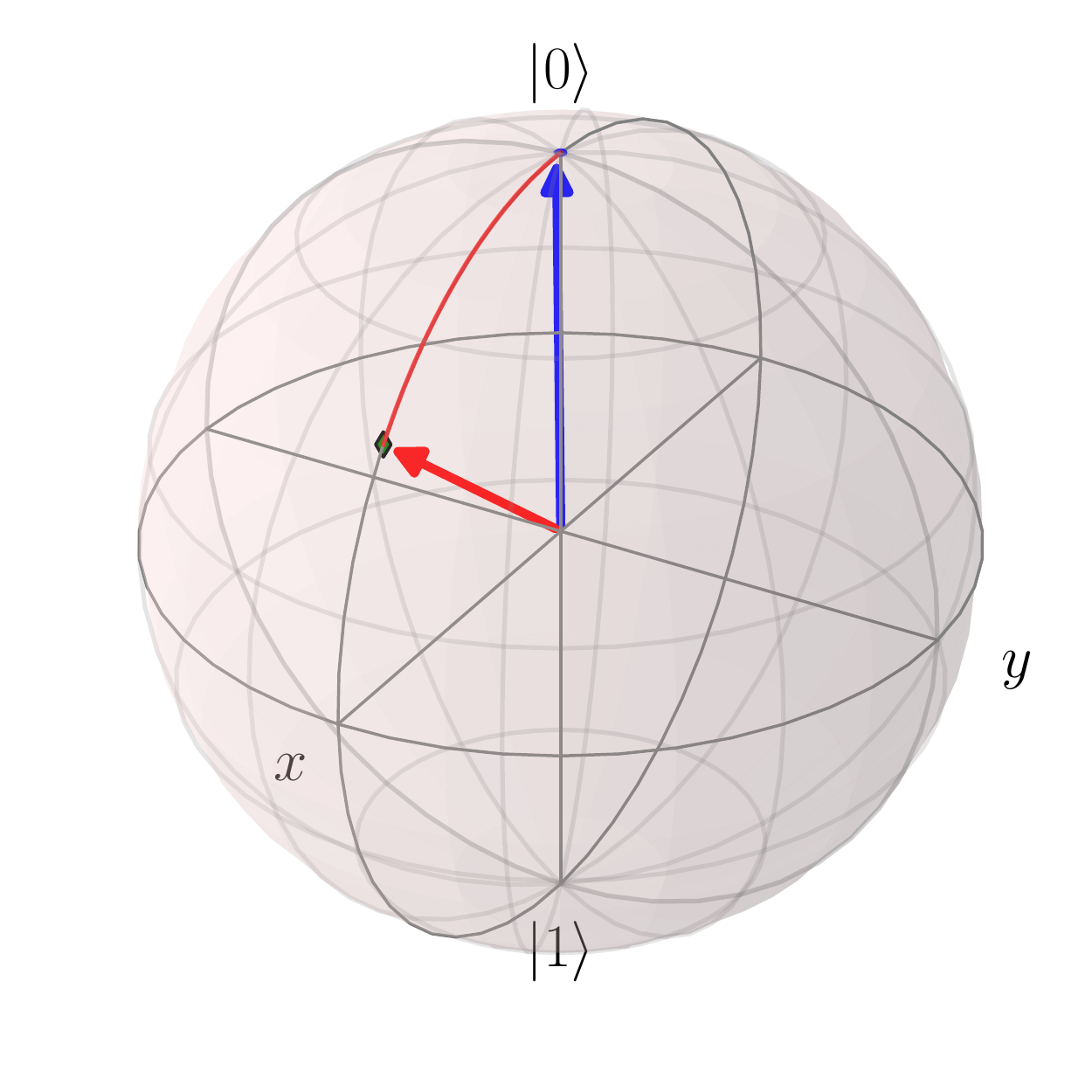} 
\caption{\label{fig:evol}
(a) $\xi = 0.5$. (b) $\xi = 0.1$. (c) $\xi = 0.01$. Time evolution of the system state (blue) and probe state (red) on the Bloch sphere, as generated by the Hamiltonian~\eqref{eq:7vfdklkln}. The strength $\xi$ of the measurement interaction (relative to the self-Hamiltonian of the system) decreases from (a) to (c). The measured system observable is $\bopvecgr{\sigma} \cdot \buvec{m}$ with $\buvec{m}=(1,1,1)$. Both system and probe are initialized in the state $\ket{0}$, and the interaction rotates the probe state about the $y$ axis. The states at the conclusion of the measurement at $t=T$ are shown as vectors. The rotation angle at $t=T$ expected for an ideal protective measurement is shown as a dot (green). }
\end{figure*} 

\subsection{Gate representation of protective measurement}

We note here that the evolution generated by the Hamiltonian~\eqref{eq:7vfdklkln} may be thought of as a controlled-rotation gate \cite{Nielsen:2000:tt}. This is so because for an ideal protective measurement, the evolution is
\begin{subequations}\label{eq:7vfdgr78ln}
\begin{align}
\ket{0}_S \ket{0}_P &\,\longrightarrow \,\ket{0}_S \op{R}_y\left(\frac{\pi}{2}  \bra{0} \bopvecgr{\sigma} \cdot \buvec{m}  \ket{0}\right) \ket{0}_P, \\
\ket{1}_S \ket{0}_P &\,\longrightarrow \,\ket{1}_S \op{R}_y\left(\frac{\pi}{2}  \bra{1} \bopvecgr{\sigma} \cdot \buvec{m}  \ket{1}\right) \ket{0}_P \notag \\ & \quad \, = \ket{1}_S \op{R}_y\left(-\frac{\pi}{2}  \bra{0} \bopvecgr{\sigma} \cdot \buvec{m}  \ket{0}\right) \ket{0}_P,
\end{align}
\end{subequations}
where $\op{R}_y(\Theta) = \E^{ -\I \Theta \op{\sigma}_y/2}=  \cos \frac{\Theta}{2} \op{I} - \I \sin \frac{\Theta}{2}  \op{\sigma}_y$ is the rotation operator for rotations by $\Theta$ around the $y$ axis on the Bloch sphere. The matrix representation of the controlled-rotation gate~\eqref{eq:7vfdgr78ln} therefore is
\begin{equation}
\begin{pmatrix}
\cos \Theta_\buvec{m} & -\sin \Theta_\buvec{m} & 0 & 0 \\
\sin \Theta_\buvec{m} & \cos \Theta_\buvec{m} &  0 & 0 \\
0 & 0 & \cos \Theta_\buvec{m} & \sin \Theta_\buvec{m} \\
0 & 0 & -\sin \Theta_\buvec{m} & \cos \Theta_\buvec{m}
\end{pmatrix},
\end{equation}
where $\Theta_\buvec{m} = \frac{\pi}{4}  \bra{0} \bopvecgr{\sigma} \cdot \buvec{m}  \ket{0}= \frac{\pi}{4}  \cos\gamma$. This gate is different from the usual controlled-rotation gates considered in quantum computation, because the state of the system qubit determines both the sign and the angle of the rotation.

\section{\label{sec:nonid-meas}Nonideal measurements}

The ideal protective measurement as represented by Eq.~\eqref{eq:vihd554ass7cf6gv} is based on the assumptions that the measurement strength $\xi$ is vanishingly small and that the self-Hamiltonian of the probe can be neglected. In Secs.~\ref{sec:noninf-meas-strength} and \ref{sec:infl-intr-probe}, we will relax these assumptions and study the resulting effect on the dynamics and relevant expectation values. We will also study the influence of interactions with an environment during the measurement (Sec.~\ref{sec:infl-inter-with}).

\subsection{\label{sec:noninf-meas-strength}Noninfinitesimal measurement strengths}

We consider the realistic case of nonideal protective measurements in which the measurement strength $\xi$ is small but, unlike in an ideal protective measurement, noninfinitesimal. In this case, the system and probe will become entangled \cite{Dass:1999:az,Schlosshauer:2014:tp}, which has several effects. At the level of the system, it results in a change of its state during the measurement \cite{Schlosshauer:2014:pm,Schlosshauer:2015:uu}. At the level of the probe, its rotation angle will be influenced, and the probe state will become partially mixed.

First, by using the exact expression~\eqref{eq:vihdgaass7cf6gv} for the final  joint state of system $S$ and probe $P$, we find that the ideal value~\eqref{eq:1} for the probe expectation value is correct to first order in $\xi$. To explore deviations from this first-order treatment, in Fig.~\ref{fig:evol} we show results of numerical calculations \cite{Johansson:2013:oo} for the evolution of the system and probe qubits during the measurement interval $t\in[0,T]$ generated by the Hamiltonian~\eqref{eq:7vfdklkln} and for different measurement strengths $\xi$, visualized on the Bloch sphere. 

We see that, as expected \cite{Schlosshauer:2014:pm,Schlosshauer:2015:uu}, the disturbance of the state of the system $S$ by the measurement decreases as $\xi$ becomes smaller. We quantify this disturbance $\mathcal{D}$ in terms of the smallest overlap between the time-evolved state $\op{\rho}_S(t)$ of $S$ with the initial state $\ket{0}$ over the course of the measurement, which we can express as 
\begin{equation}
\mathcal{D} = 1-\min_{0 \le t \le T} \text{Tr}\left[\op{\rho}_S(t) \op{\sigma}_z \right].
\end{equation}
For $\xi = 0.5$, the state disturbance of the state of $S$ is significant ($\mathcal{D}=49\%$) and the purity of the final states of $S$ and $P$ is only 0.82, indicating substantial entanglement between system and probe. For $\xi = 0.1$, the state disturbance for $S$ has become very small ($\mathcal{D}=3\%$), and the final states of $S$ and $P$ retain nearly complete (0.99) purity.  Thus, the measurement can be considered protective. For $\xi = 0.01$, the state disturbance of $S$ is negligibly small. 

Looking at the rotation of the probe qubit $P$ as shown in Fig.~\ref{fig:evol}, we first note that it is always in the $xz$ plane, which is expected since the probe evolution is solely due to the $\op{\sigma}_y$ term in Eq.~\eqref{eq:7vfdklkln}. We also see that, as the measurement strength decreases, the total rotation angle of the probe qubit quickly approaches the value for an ideal protective measurement (as indicated by a dot in Fig.~\ref{fig:evol}). Recall that the rotation angle can be obtained from the expectation value $\langle\op{\sigma}_x(t)\rangle$ for the probe (see Sec.~\ref{sec:probe-readout}). We can then compare this expectation value at the conclusion of the measurement ($t=T$) to the value $\sin\left(\frac{\pi}{2}\bra{0} \bopvecgr{\sigma} \cdot \buvec{m}  \ket{0}\right)$ [see Eq.~\eqref{eq:1}] expected for an ideal protective measurement. For $\xi = 0.5$, the expectation value $\langle\op{\sigma}_x(T)\rangle$ differs substantially (22\%) from the ideal value, indicating that in this regime the pointer shift (i.e., the rotation angle) does not yet faithfully reproduce the ideal value. For $\xi = 0.1$, the difference between actual and ideal expectation values is only 1.5\%. 

\subsection{\label{sec:repeated}Repeated measurements}

As discussed in Sec.~\ref{sec:probe-readout}, probe readout requires measurement of an expectation value. In practice, such a measurement may be realized through $N$ repeated cycles consisting of probe preparation, system--probe interaction, and probe measurement. Since the state of the system at the end of the $n$th measurement becomes the initial state of the system for the $(n+1)$th measurement, the disturbance imparted on the system becomes propagated through the series of consecutive measurements. This raises the question of how the performance of the scheme will be affected by such a series of protective measurements. 

\begin{figure}
(a) \hspace{3.6cm} (b)
  
\includegraphics[width=1.65in]{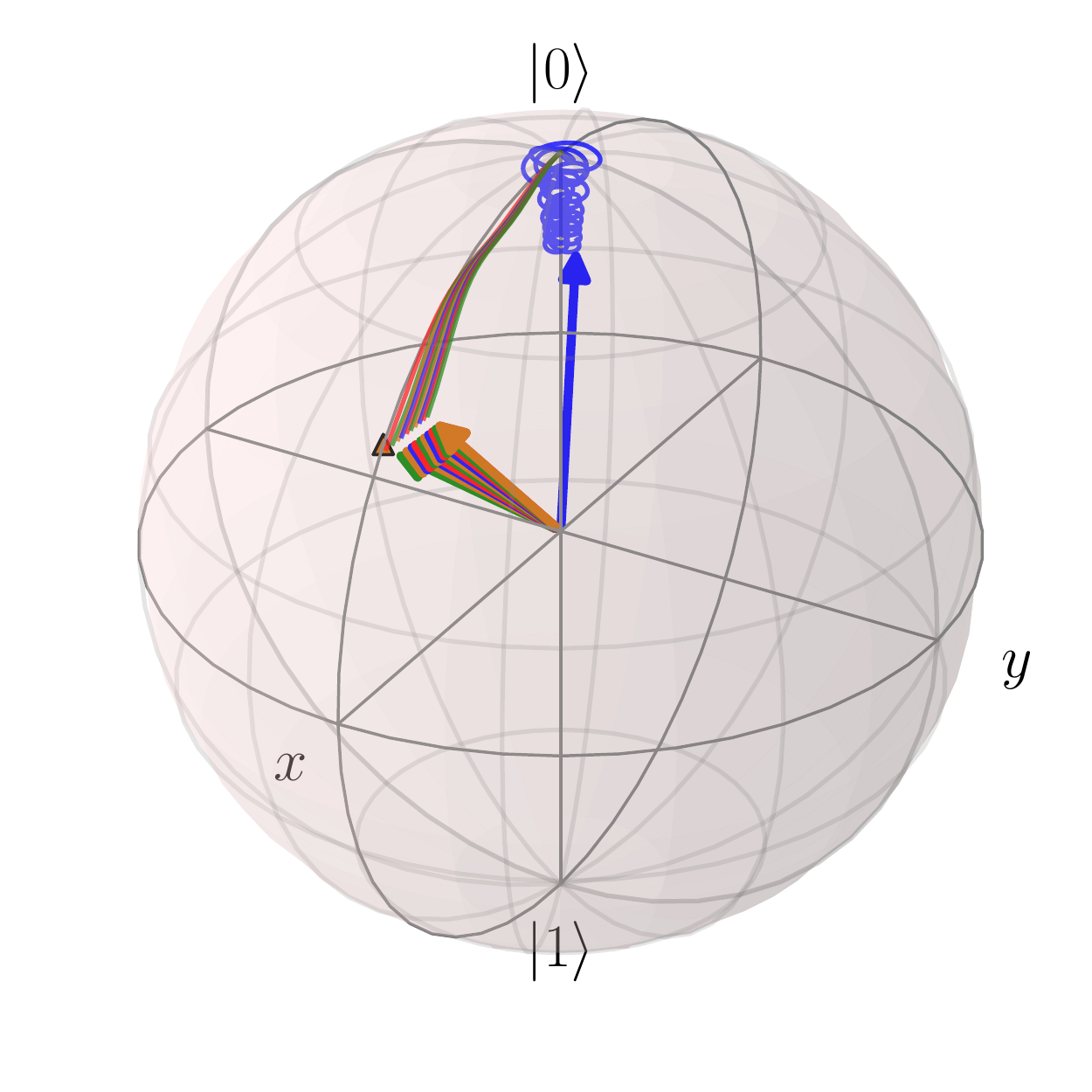} \includegraphics[width=1.65in]{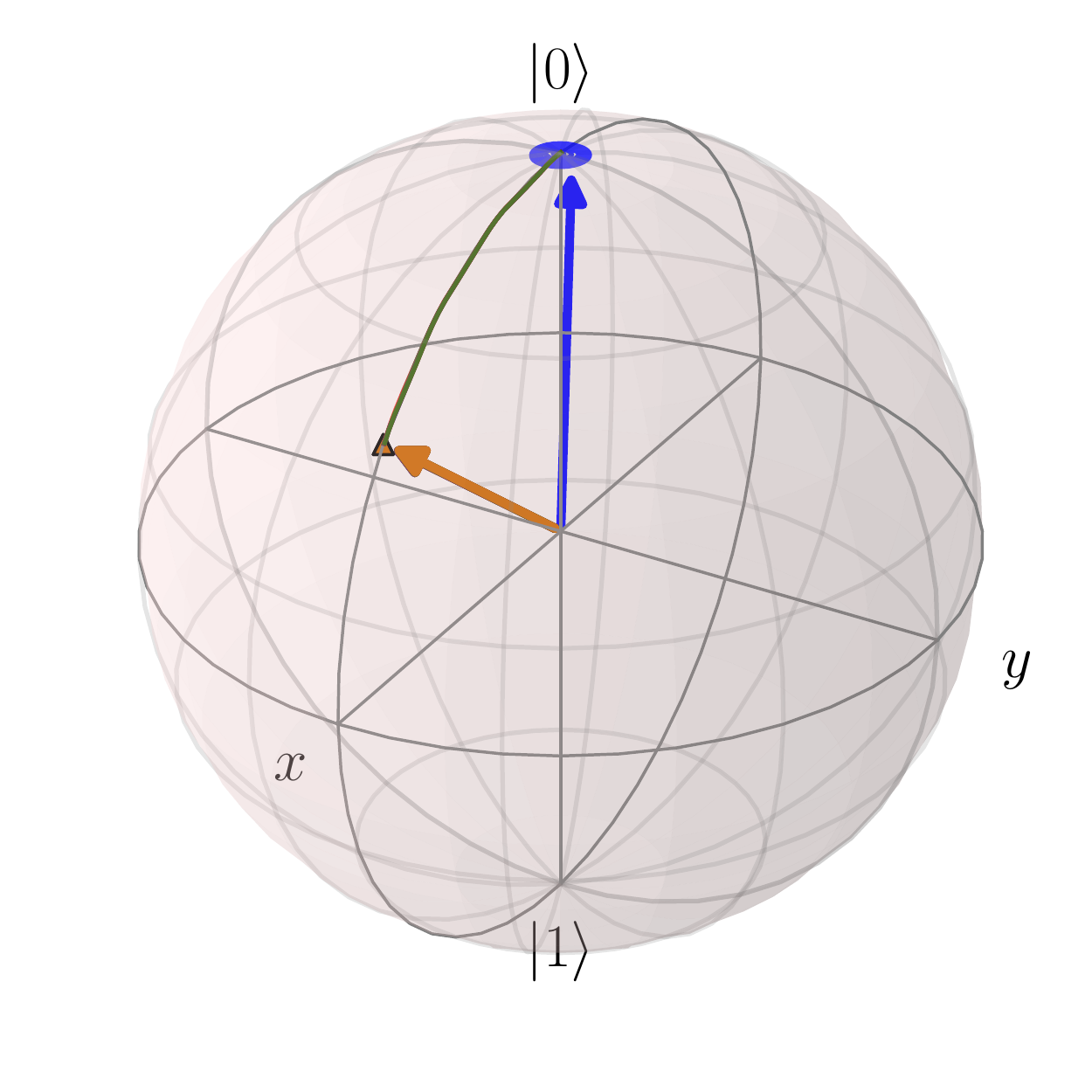} 
\caption{\label{fig:repeat}(a) $\xi = 0.1$. (b) $\xi = 0.05$. Time evolution of the system state and probe states during a series of ten consecutive measurements, shown for two different measurement strengths $\xi$. At the start of each measurement, the probe is initialized in the same state $\ket{0}$, while the state of the system evolves along the chain of measurements, each of which is described by the Hamiltonian~\eqref{eq:7vfdklkln}. The blue vector near the $z$ axis of the Bloch sphere shows the final state of the system (at $t=10T$). The remaining vectors represent the probe states at the conclusion of each of the ten individual measurements (for $\xi = 0.05$, these vectors essentially coincide and thus appear as a single vector).}
\end{figure} 

To explore this issue, we study the time evolution of the system and probe states in the course of $N=10$ consecutive measurements. Each measurement is of duration $T$, and the probe is initialized in the same state  (the eigenstate $\ket{0}$ of $\op{\sigma}_z$, as before) at the start of each measurement. The results are shown in Fig.~\ref{fig:repeat}, where we also display the probe states at the end of each of the ten measurements. If the measurement interaction is only moderately weak ($\xi=0.1$), then, as expected, the accumulated state disturbance of the system will be significant, around 25\%, with a purity of 0.88. The final individual probe states are seen to differ somewhat, since each of them corresponds to the measurement of a slightly different system state. As expected, the difference between actual and ideal values for the expectation value quantifying the probe rotation increases along the chain of measurements, because the system increasingly departs from its initial state as the number of measurements increases. As a worst-case estimate, we take the probe state obtained from the final measurement to estimate the difference between actual and ideal values for the probe rotation, which gives 24\%. When we instead average over all ten probe states, the difference is 14\%.

The detrimental influence of multiple measurements on the quality of the protective measurement rapidly diminishes as the interaction is made weaker. For $\xi=0.05$ [shown in Fig.~\ref{fig:repeat}(b)], the cumulative disturbance of the state of the system is reduced to only 1.6\%. The Bloch vectors of all final probe states are seen to coincide, with a difference between ideal and actual values for the probe rotation of only 0.7\% using the worst-case estimate. For $\xi=0.01$, no discernible difference is observed, in terms of state disturbance and faithfulness of the probe rotation, for the series of ten protective measurements when compared to a single measurement. These results suggest that the need for multiple system--probe interactions and subsequent probe readouts does not pose a significant challenge to the protective-measurement scheme based on a qubit probe.

\subsection{\label{sec:infl-intr-probe}Intrinsic probe dynamics}

\begin{figure*}
(a)  \hspace{3.7cm} (b)  \hspace{3.7cm}  (c)

\includegraphics[width=1.65in]{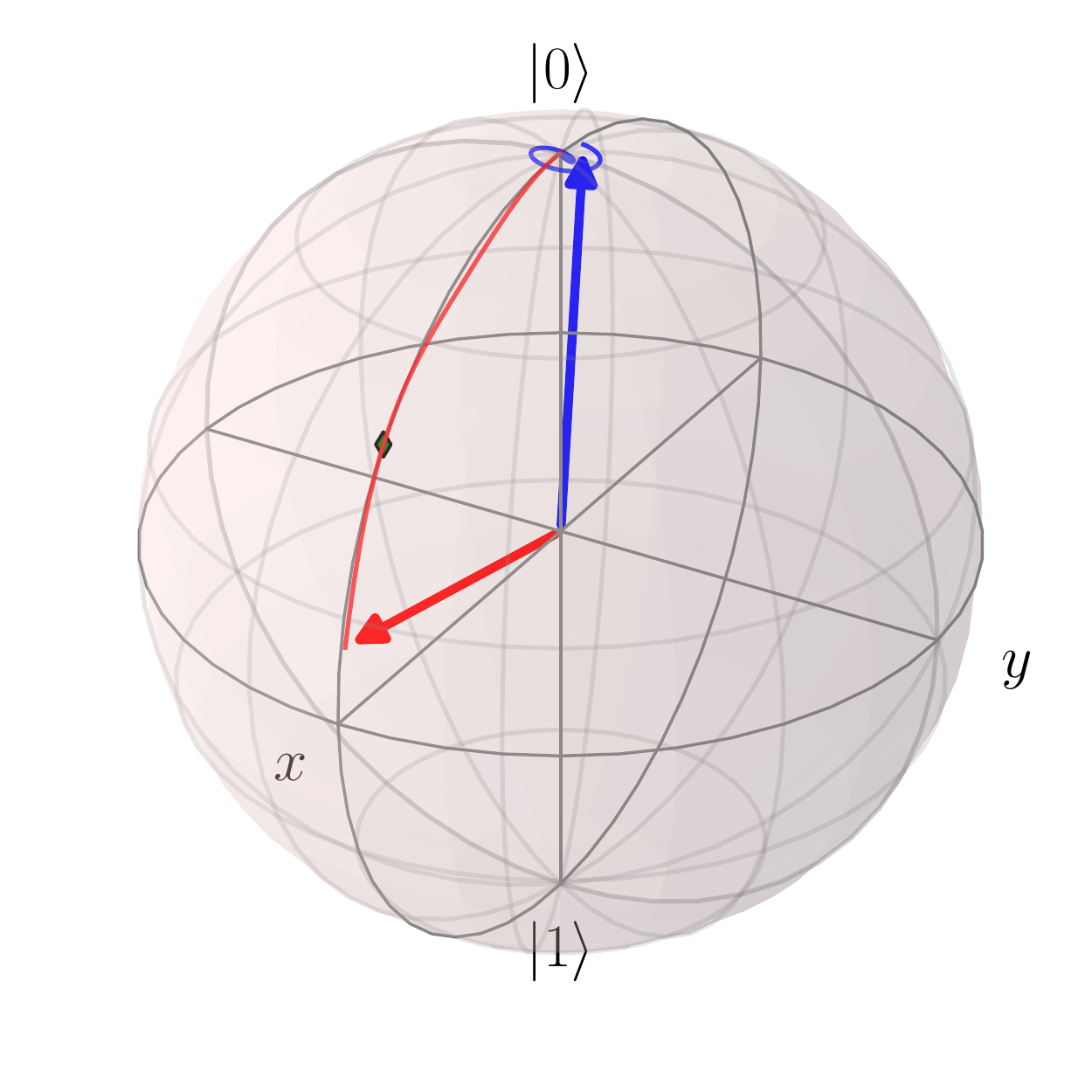} \includegraphics[width=1.65in]{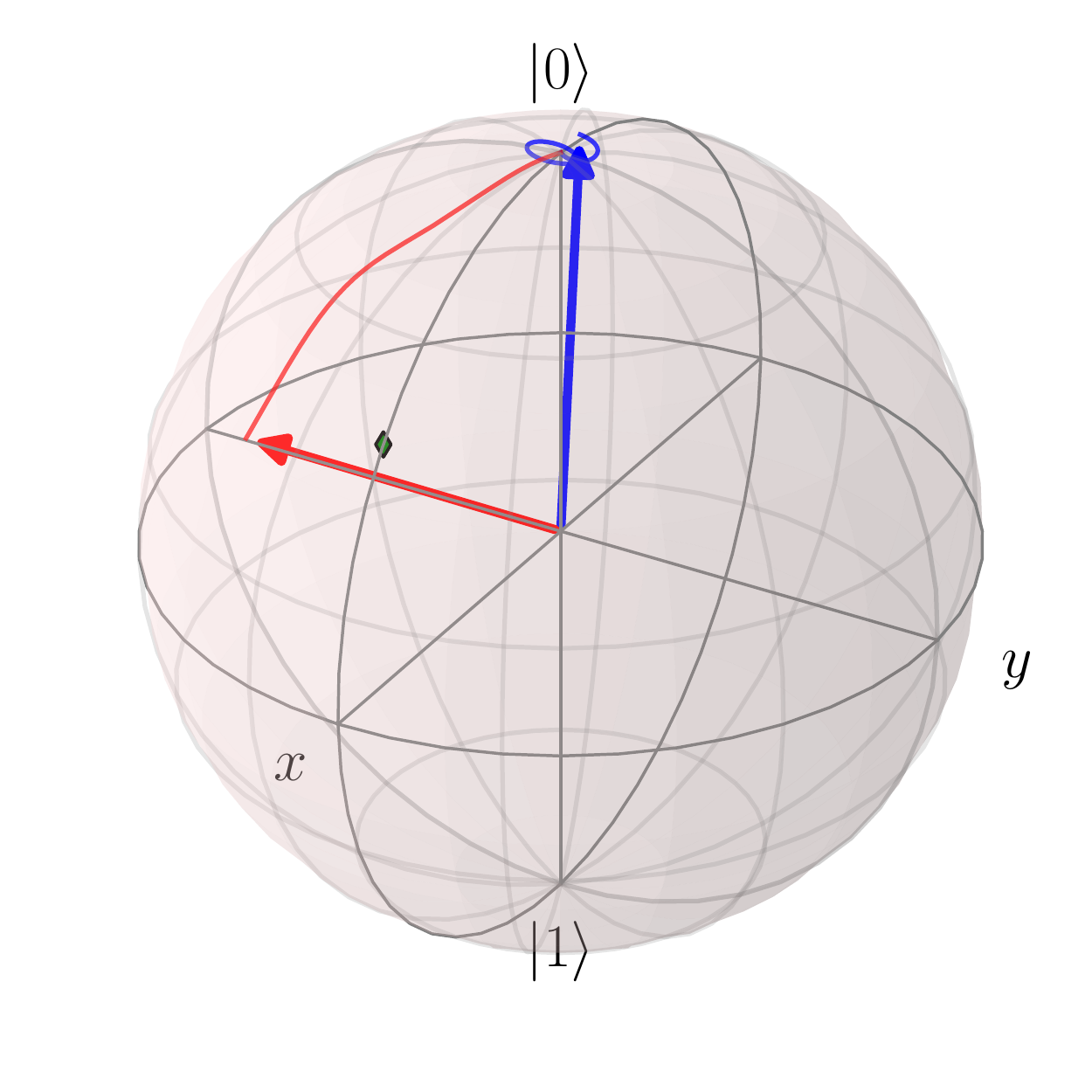} \includegraphics[width=1.65in]{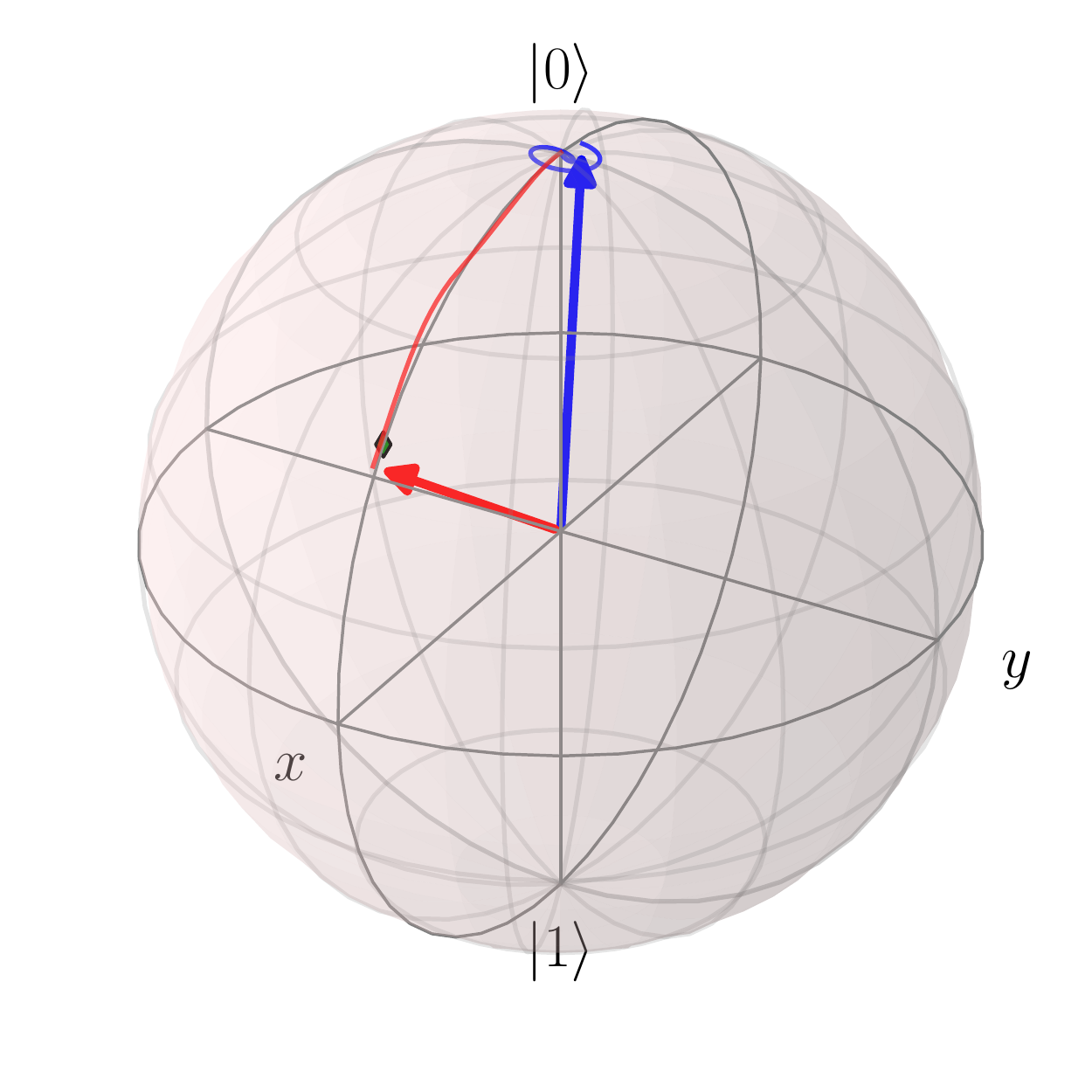} 
\caption{\label{fig:intdyn}
(a) $\op{\sigma}_y$, $\delta_P=0.3$. (b) $\op{\sigma}_x$, $\delta_P=0.3$. (c) $\op{\sigma}_x+\op{\sigma}_y+\op{\sigma}_z$, $\delta_P=0.05$. 
Time evolution of the system state (blue) and probe state (red) in the presence of intrinsic probe dynamics generated by a self-Hamiltonian $\op{H}_P = \frac{\hbar\pi}{4T}\delta_P (\bopvecgr{\sigma} \cdot \buvec{r})$. The Pauli operator $(\bopvecgr{\sigma} \cdot \buvec{r})$ and the parameter $\delta_P$ quantifying the strength of $\op{H}_P$ relative to the interaction Hamiltonian $\op{H}_m$ are shown for each panel. The measurement strength is $\xi = 0.1$, and the measured system observable and initial states are the same as in Fig.~\ref{fig:evol}.}
\end{figure*} 

So far, we have neglected the self-Hamiltonian $\op{H}_P$ of the probe. This approximation is common to most considerations of protective measurement (but see Ref.~\cite{Dass:1999:az} for some general results on the influence of a nonzero probe Hamiltonian). To explore the influence of intrinsic probe dynamics in our model, we add to the Hamiltonian~\eqref{eq:7vfdklkln} a generic probe self-Hamiltonian $\op{H}_P =\frac{1}{2}\hbar \omega_P (\bopvecgr{\sigma} \cdot \buvec{r})$. This Hamiltonian will contribute a rotation of the probe state around the $\buvec{r}$ axis. Thus, the evolution of the probe will now consist of a rotation around a new axis given by a linear  combination of the $y$ and $\buvec{r}$ axes. 

Clearly, for the probe to faithfully encode the desired expectation value of the system, the contribution of the $\buvec{r}$ axis to the net axis should in general be small, such that the probe dynamics are dominated by the interaction with the measured qubit system $S$ (indeed, this is a sensible requirement for any quantum system designated to act as a measuring device). To ensure that this condition holds for any choice of $\buvec{r}$, one therefore needs to require that $\op{H}_P$ be small compared with the interaction Hamiltonian $\op{H}_m$. Since the strength of $\op{H}_m$ is given by $\hbar\pi/4T$ [compare Eq.~\eqref{eq:7vfdklkln}], we quantify the strength of $\op{H}_P$ relative to $\op{H}_m$ by writing $\omega_P = \frac{\pi}{2T}\delta_P$, where $\delta_P$ is a dimensionless parameter that represents the ratio of the  strength of $\op{H}_P$ to the strength of $\op{H}_m$. 

The particular effect of $\op{H}_P$ on the probe rotation will depend on the choice of the $\buvec{r}$ axis. If $\buvec{r}=\buvec{y}$, i.e., if the probe Hamiltonian $\op{H}_P$ is proportional to $\op{\sigma}_y$, then this Hamiltonian will leave the probe state in the $xz$ plane but add a constant $\frac{1}{2} \omega_PT = \frac{\pi}{4} \delta_P$ to the rotation angle. This is illustrated in Fig.~\ref{fig:intdyn}(a) for a moderately weak probe Hamiltonian ($\delta_P=0.3$). The overshoot of the rotation is clearly seen, and we find that the expectation value $\langle\op{\sigma}_x(T)\rangle$ quantifying the rotation angle differs by 22\% from the value~\eqref{eq:1} that would be obtained for an ideal protective measurement with $\op{H}_P=0$.

Next, let us consider the situation in which the probe Hamiltonian $\op{H}_P$ is proportional to $\op{\sigma}_x$. The net rotation axis is now in the $xy$ plane, and the probe state will be rotated out of the $xz$ plane [see Fig.~\ref{fig:intdyn}(b)]. However, because the change in the rotation axis due to $\op{H}_P$ is perpendicular to the $\op{\sigma}_y$ rotation axis for the system--probe interaction, the influence on the projection of the Bloch vector on the $x$ axis [as given by $\langle\op{\sigma}_x(T)\rangle$] can be less dramatic as in the previous case of $\buvec{r}=\buvec{y}$. Indeed, for the example shown in Fig.~\ref{fig:intdyn}(b), the difference between ideal and actual values of $\langle\op{\sigma}_x(T)\rangle$ is only 5\% at $\delta_P=0.3$. 
Finally, Fig.~\ref{fig:intdyn}(c) shows the evolution when the probe Hamiltonian $\op{H}_P$ is proportional to $\op{\sigma}_x+\op{\sigma}_y+\op{\sigma}_z$ [i.e., $\buvec{r} = (1,1,1)/\sqrt{3}$] and its strength is reduced to $\delta_P=0.05$. As expected, now the probe rotation is only insignificantly influenced by the intrinsic probe dynamics, and the difference between ideal and actual values of $\langle\op{\sigma}_x(T)\rangle$ is 2\%. One would not expect the addition of a self-Hamiltonian for the probe to affect the purity of the final system and probe states, because no entanglement is created by this Hamiltonian. We have explicitly confirmed this expectation by calculating the state purities in each of the cases shown in Fig.~\ref{fig:intdyn} and finding the same purity value (0.99) as in the absence of $\op{H}_P$.

\begin{figure}
(a) $\op{\sigma}_y$ \hspace{3.6cm} (b)

\includegraphics[width=1.65in]{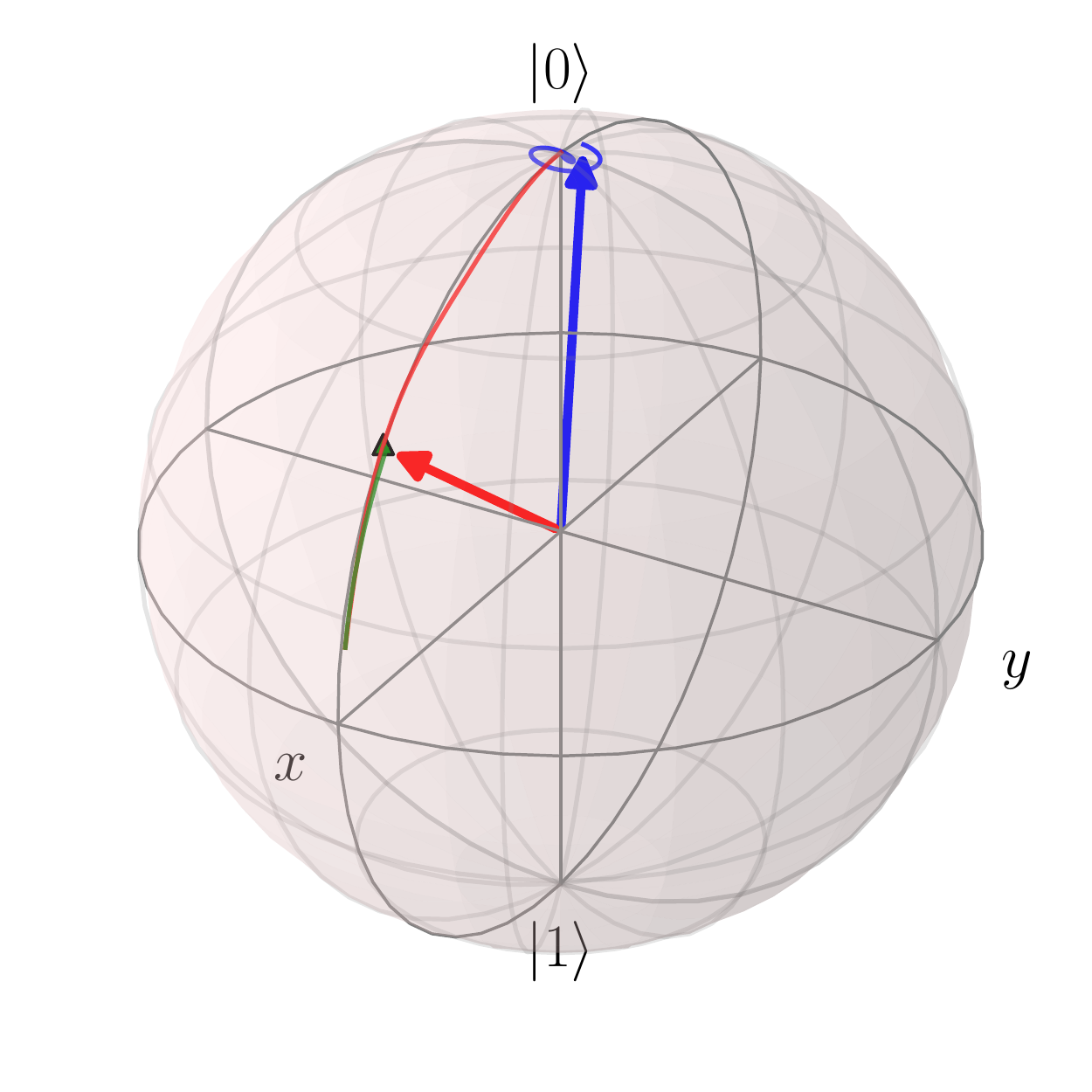} \includegraphics[width=1.65in]{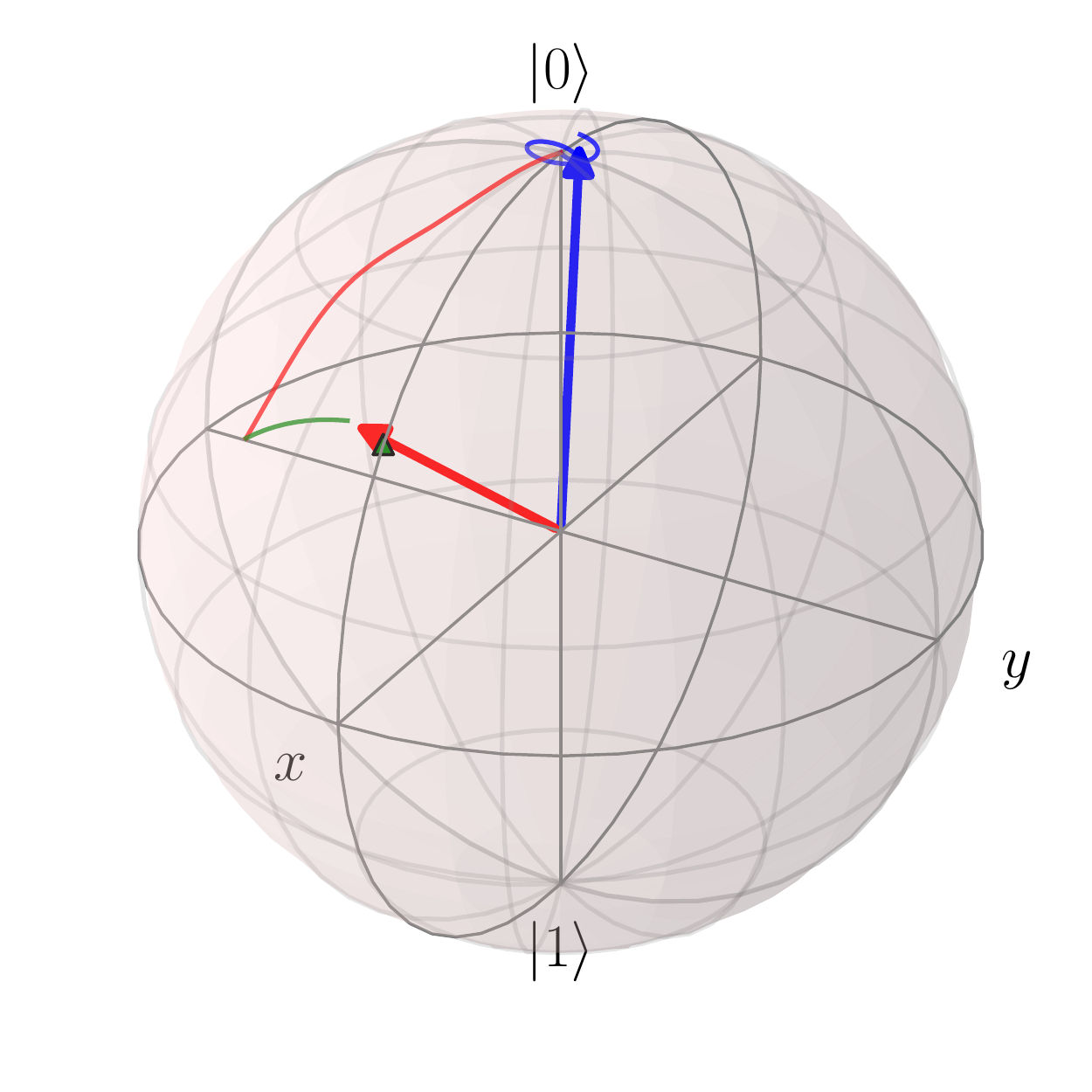} 
\caption{\label{fig:correct}(a) $\op{\sigma}_y$. (b) $\op{\sigma}_x$. Effect of an applied counter-rotation to mitigate the effect of the intrinsic probe dynamics.  The evolution of the probe state during the measurement is shown in red, and the subsequent correcting evolution due to the applied counter-rotation is shown in green. The evolution of the system state (blue) is the same as in Fig.~\ref{fig:intdyn}(a) and shown for reference. (a) When the intrinsic probe dynamics produce a rotation around the same  axis as the measurement interaction, the effect of the intrinsic dynamics is reversed. (b) When the intrinsic rotation is instead around the $x$ axis, the expectation value $\langle \op{\sigma}_x \rangle$ remains unchanged by the counter-rotation. The parameter values are $\delta_P=0.3$ and $\xi = 0.1$, and the measured system observable and initial states are the same as in Fig.~\ref{fig:evol}.}
\end{figure} 

We have seen that in the case where the axis $\buvec{r}$ for the intrinsic rotation coincides with the axis for the probe rotation due to the measurement interaction (here the $y$ axis), the effect of the probe Hamiltonian is to produce a simple overshoot of the probe state [as shown in Fig.~\ref{fig:intdyn}(a)], i.e., the effect is to modify the rotation angle but not the rotation axis. Such an overshoot of the probe is easily corrected by applying a counter-rotation to the probe qubit after its interaction with the system. Since we can take the probe Hamiltonian to be known, the relevant parameters ($\omega_P$ and $T$, or equivalently $\delta_P$) needed to choose the compensating rotation angle will also be known. This strategy is shown in Fig.~\ref{fig:correct}(a). The final state of the probe now correctly indicates the desired expectation value (2\% difference to the ideal value). This correction strategy does not work adequately, however, when the axes defining the measurement rotation and the intrinsic rotation of the probe are different. The extreme case is that of an intrinsic rotation around the $x$ axis [see Fig.~\ref{fig:correct}(b)]. Since the counter-rotation around $x$ will preserve the value of $\langle \op{\sigma}_x \rangle$, it will not improve the fidelity of the measurement result. These results suggest that in cases where the intrinsic dynamics of the probe during the measurement are significant, the measurement interaction should be chosen such that the probe rotation is around the same axis as the rotation produced by the intrinsic dynamics.

\subsection{\label{sec:infl-inter-with}Interactions with an environment}

In the Hamiltonian~\eqref{eq:7vfdklkln}, the only interaction of the system qubit is with the probe qubit. In realistic physical settings (such as the ion-trap experiment proposed in Sec.~\ref{sec:prop-exper-impl}), both qubits may also be subject to decoherence and dissipation due to interactions with their environment (noise processes give rise to phenomenologically similar effects) \cite{Breuer:2002:oq,*Schlosshauer:2019:qd}. We will now include such environmental effects in the Hamiltonian~\eqref{eq:7vfdklkln} by coupling both qubits to bosonic baths. We model the resulting dynamics in terms of a Lindblad master equation \cite{Breuer:2002:oq} for the joint density operator $\op{\rho}_{SP}(t)$ of the system $S$ and the probe $P$, 
\begin{align}\label{eq:lindbladc}
\frac{\partial}{\partial t} \op{\rho}_{SP} (t) &= - \frac{\I}{\hbar} \left[ \op{H}'_S, \op{\rho}_{SP} (t) \right] \notag\\ &\quad - \frac{1}{2} \sum_{k=S,P} \kappa_k \left[ \op{L}_k, \left[ \op{L}_k, \op{\rho}_{SP} (t) \right]\right],
\end{align}
where $\op{H}'$ is the Hamiltonian~\eqref{eq:7vfdklkln} renormalized by the environment, $\op{L}_S =  (\bopvecgr{\sigma} \cdot \buvec{e}_S)\otimes \op{I}$ and $\op{L}_P = \op{I} \otimes  (\bopvecgr{\sigma} \cdot \buvec{e}_P)$ are the Lindblad operators representing the coupling of $S$ and $P$ to the environment, and $\kappa_S$ and $\kappa_P$ are the corresponding rates.

\begin{figure}
\includegraphics[width=1.65in]{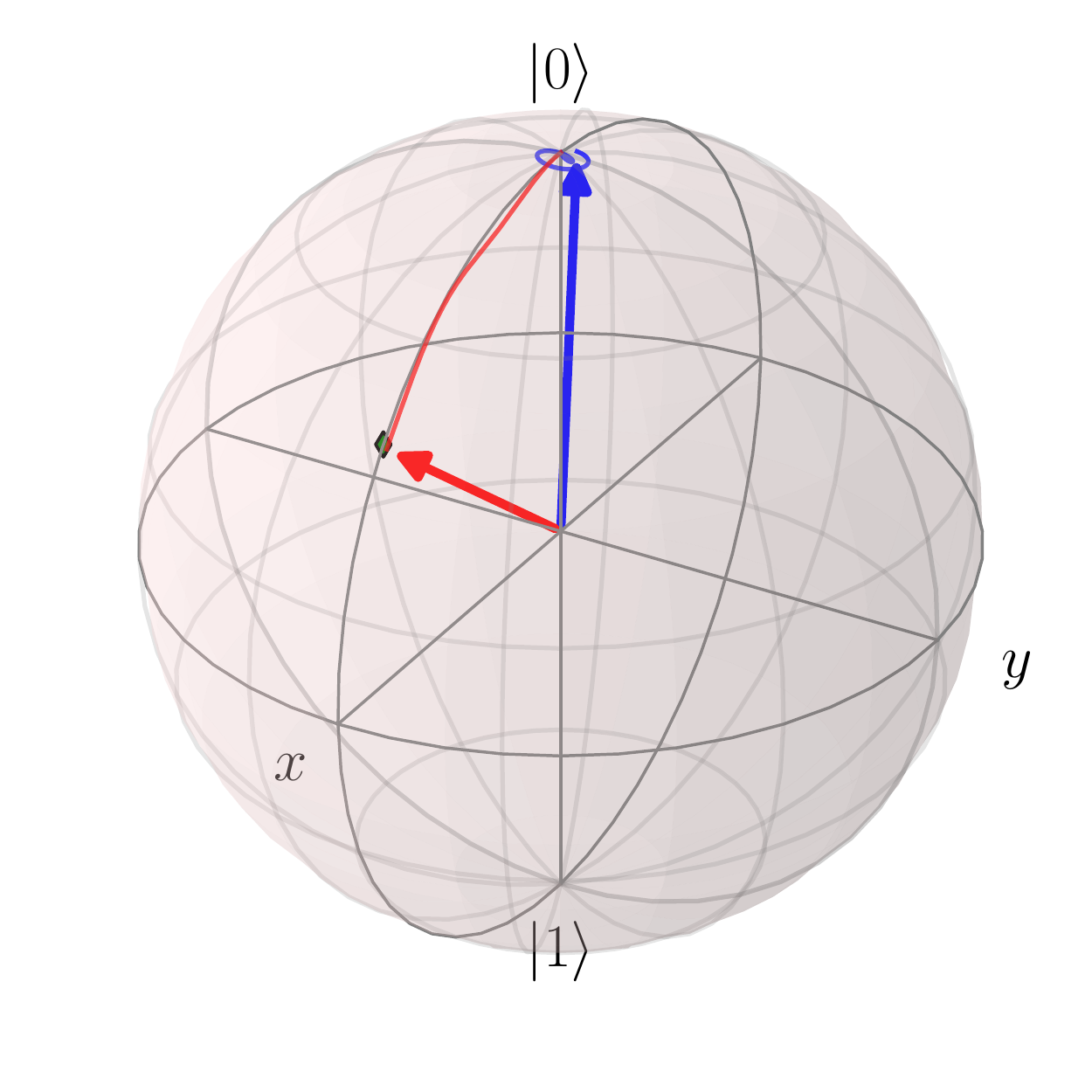} 
\caption{\label{fig:envsz}Time evolution of the system state (blue) and probe state (red) when the system is coupled to a bosonic environment via the $\op{\sigma}_z$ coordinate. The dynamics are modeled in terms of the Lindblad master equation~\eqref{eq:lindbladc}. The rate is $\kappa_S=0.02$, the measurement strength is $\xi = 0.1$, and the measured system observable and initial states are the same as in Fig.~\ref{fig:evol}.}
\end{figure} 

\begin{figure*}
(a)  \hspace{3.7cm} (b)  \hspace{3.7cm}  (c)

\includegraphics[width=1.65in]{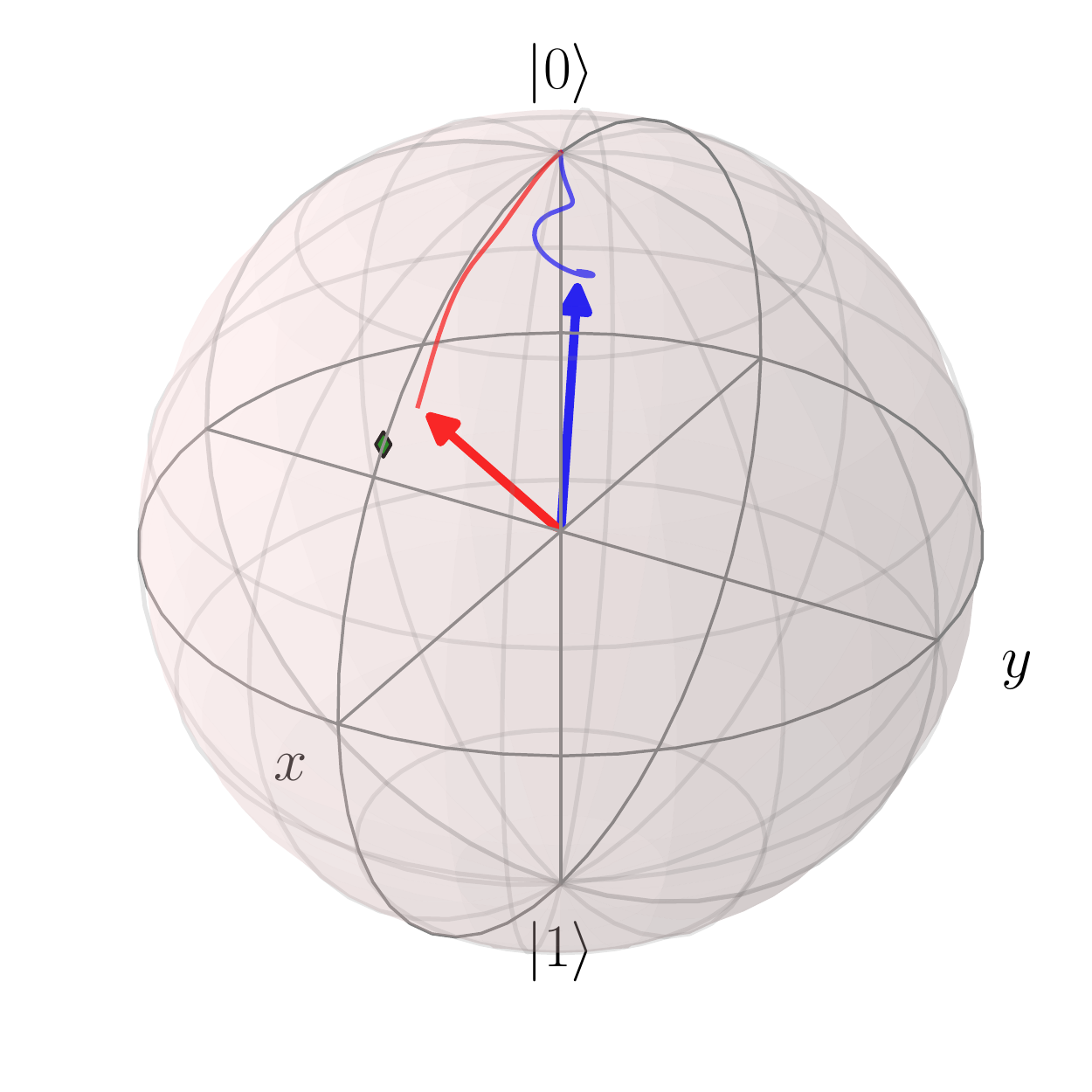} \includegraphics[width=1.65in]{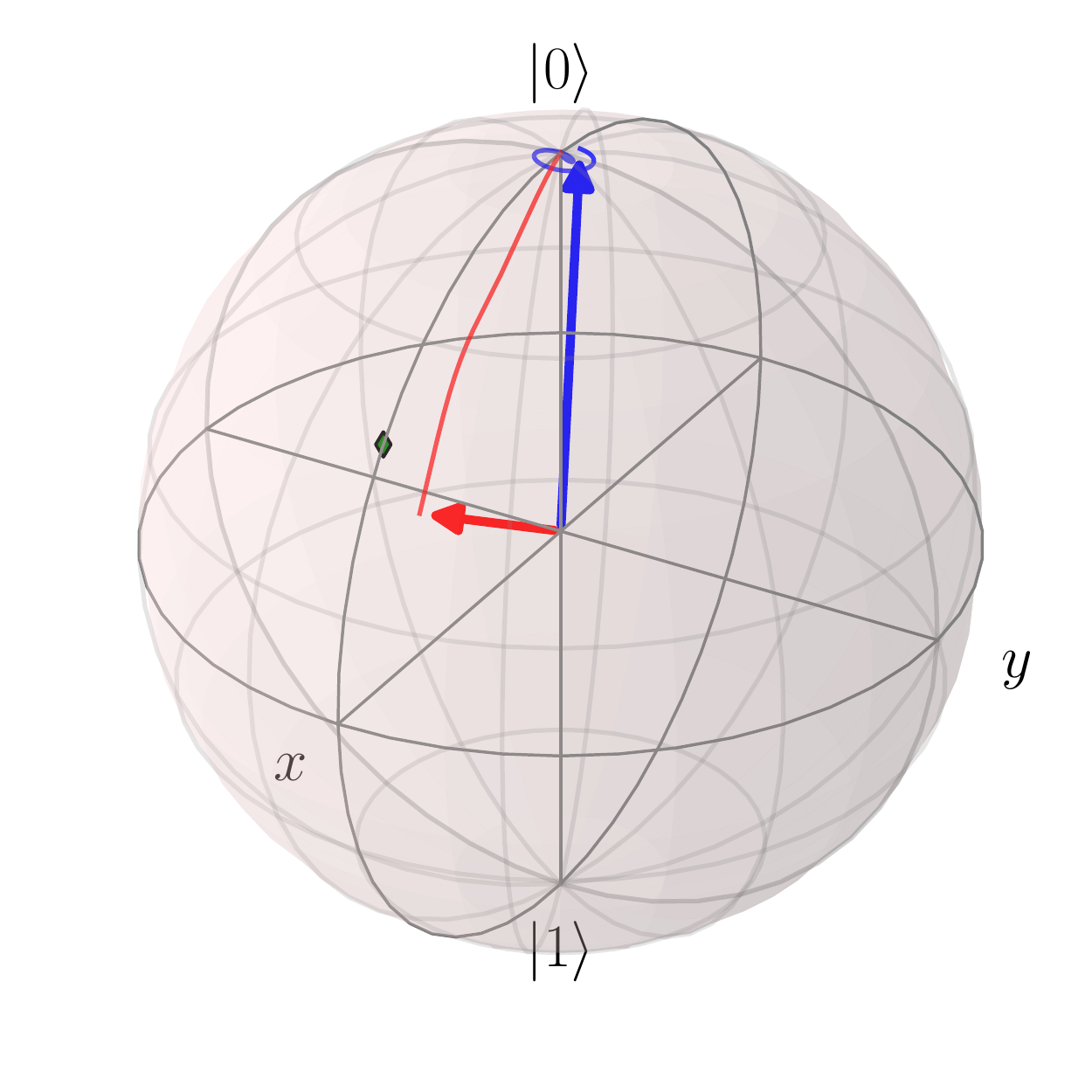} \includegraphics[width=1.65in]{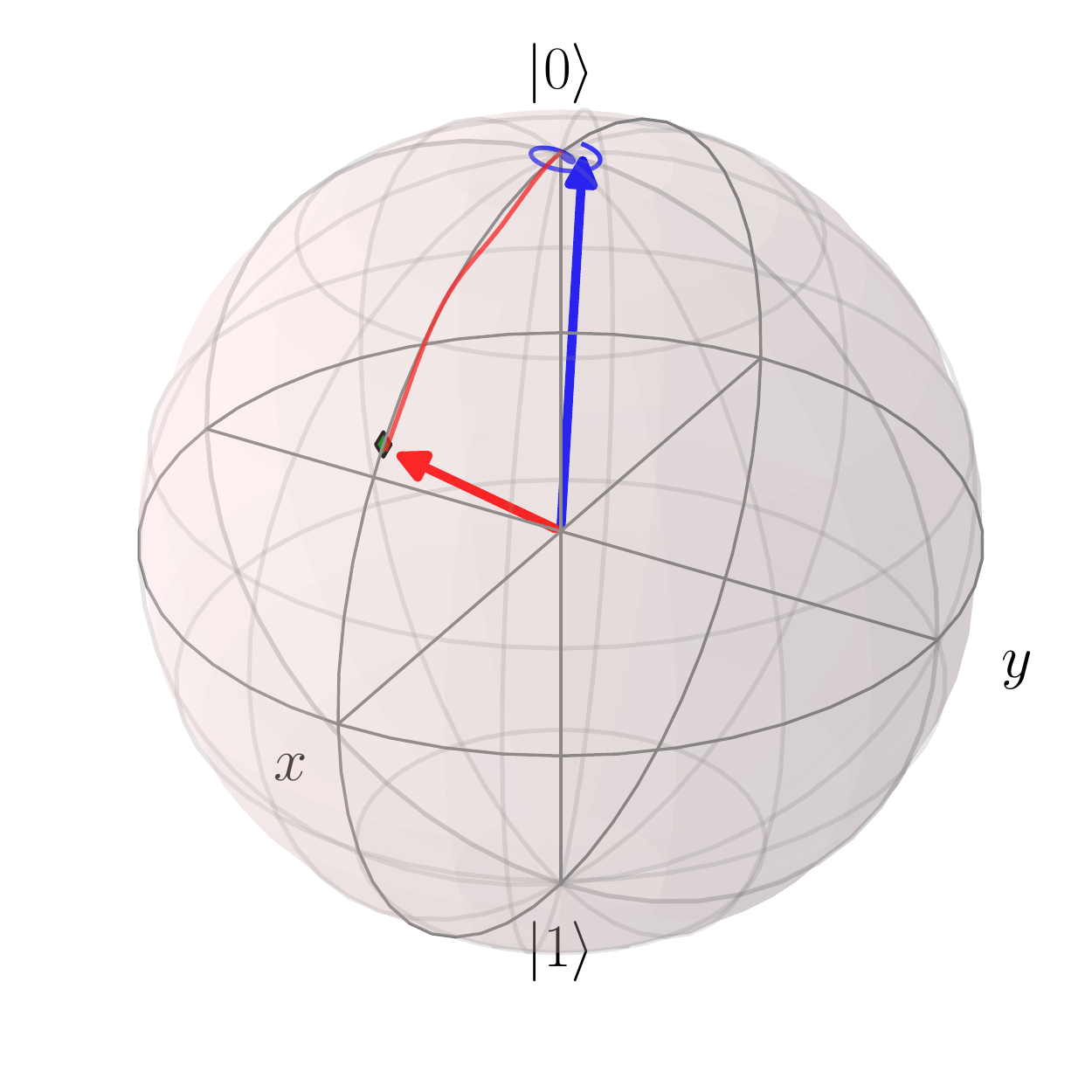} 
\caption{\label{fig:env}
(a) $\kappa_S=0.02$, $\kappa_P=0$. (b) $\kappa_S=0$, $\kappa_P=0.02$. (c) $\kappa_S=\kappa_P=0.02$. Time evolution of the system state (blue) and probe state (red) in the presence of an environment. System and probe are coupled to bosonic baths through their $\op{\sigma}_x$ coordinates, and the dynamics are modeled in terms of the Lindblad master equation~\eqref{eq:lindbladc}. The chosen rates $\kappa_S$ and $\kappa_P$ of the environment-induced processes are shown in each panel.  The measurement strength is $\xi = 0.1$, and the measured system observable and initial states are the same as in Fig.~\ref{fig:evol}. (a) Evolution when only the system couples to the environment. (b) Evolution when only the probe couples to the environment. (c) Evolution when both the system and the probe couple to the environment.}
\end{figure*} 

Since the system $S$ remains throughout the measurement close to the eigenstate $\ket{0}$ by virtue of its dominant self-Hamiltonian $\op{H}_S$, a coupling to the environment via its $\op{\sigma}_z$ coordinate is expected to have little effect on the evolution. This is illustrated in Fig.~\ref{fig:envsz}. Comparison with Fig.~\ref{fig:evol}(b) for the evolution in the absence of an environment indicates that the environment has indeed no significant influence, and no discernible decrease in purity of the final system and probe states is observed. 

To produce an appreciable effect of the environment, let us now choose a coupling for both system $S$ and probe $P$ via their $\op{\sigma}_x$ coordinates. The resulting time evolution of the system and probe states is shown in Fig.~\ref{fig:env}. In Fig.~\ref{fig:env}(a), only the system $S$ couples to the environment. The Bloch vector representing the state of $S$ remains close to the $z$ axis but is substantially shortened in length, indicating an incoherent mixture (purity 0.83) of $\ket{0}$ and $\ket{1}$ with a sizable probability of finding the system in the state $\ket{1}$. This behavior is expected, since the coupling to the environment via $\op{\sigma}_x$ induces transitions between $\ket{0}$ and $\ket{1}$. The state disturbance is 35\%, a significant impact on the protective measurement given its goal of leaving the initial state of the system approximately unchanged. Moreover, Fig.~\ref{fig:env}(a) shows that although the probe does not couple directly to the environment, its rotation angle is also affected. This, too, is expected, because the evolution of the probe is entirely governed by its coupling to the system qubit $S$ interacting with the environment, and probe rotation at any instant depends on the state of $S$. Thus, the environment-induced changes of the state of $S$ translate into changes in the probe evolution, in our example resulting in an 19\% difference between actual and ideal values of $\langle\op{\sigma}_x(T)\rangle$. A small decrease in purity (0.95) of the probe state is also observed. This is reasonable, since the probe becomes entangled with a system that is in turn entangled with the environment, leading to an overall increase in the amount of entanglement of the probe.

Figure~\ref{fig:env}(b) shows the evolution if only the probe interacts with the environment. The influence of the environment on the evolution of the probe is clearly seen. The rotation remains in the $xz$ plane, but the shortening of the Bloch vector indicates that the probe state becomes appreciably mixed (purity 0.87) due to the entanglement with the environment. The difference between actual and ideal values of $\langle\op{\sigma}_x(T)\rangle$ is 19\%. We also see that the state of the system $S$ is not affected by the environment. This is expected, because the coupling of $S$ to $P$ is weak compared to the intrinsic evolution generated by the self-Hamiltonian of $S$.

Finally, Fig.~\ref{fig:env}(c) shows the evolution when both the system and the probe couple to the environment. Now the probe state is doubly affected, both by the direct coupling to its own environment and by the indirect coupling to the environment of the open system $S$. Accordingly, the Bloch vector of the probe state is further shortened compared to Fig.~\ref{fig:env}(b), indicating an increase in mixedness (purity 0.83), and the difference between actual and ideal values of $\langle\op{\sigma}_x(T)\rangle$ rises to 32\%. 

\section{\label{sec:prop-exper-impl}Proposed experimental implementation}

The protective-measurement model described in this paper can be experimentally realized with trapped ions using existing technology. All the necessary components, including state preparation, implementation of the relevant single- and two-qubit Hamiltonians, and final-state readout, are already part of existing ion-trap experiments used for quantum computation \cite{Bruzewicz:2019:aa} and quantum simulations of  spin systems \cite{Milburn:2000:az,Porras:2004:tt,Lin:2011:aa,Korenblit:2012:zz,Jurcevic:2014:uu,Hayes:2014:kk,Smith2016:im,Monroe:2019:za}. In such experiments, the qubit states $\ket{0}$ and $\ket{1}$ (eigenstates of the $\op{\sigma}_z$ operator) are formed by two internal electronic levels of an atomic ion. Preparation of the qubit state through optical pumping is accomplished within a few microseconds and achieves purities in excess of 99.9\% \cite{Harty:2014:rr}. State-dependent fluorescence  can be used to measure the final qubit states with efficiencies $> 99\%$ \cite{Noek:2013:uu,Harty:2014:rr}. The protective-measurement Hamiltonian~\eqref{eq:vdkkjl7} requires implementation of a protection term $\op{\sigma}_z$ and an interaction term $(\bopvecgr{\sigma} \cdot \buvec{m}) \otimes (\bopvecgr{\sigma} \cdot \buvec{n})$, with the protection term dominant. Both terms, with adjustable relative strengths, can be realized by the simultaneous application of suitable external laser fields that couple qubit levels either directly or via the phonon modes that describe the collective vibrational motion of the trapped ions \cite{Sorensen:1999:za,Lee:2005:kk,Blatt:2008:uu,Soderberg:2010:za,Bruzewicz:2019:aa,Monroe:2019:za}.  These are precisely the interactions used in ion traps for implementing quantum gates \cite{Bruzewicz:2019:aa} and for simulating the quantum dynamics of spin systems subject to magnetic fields \cite{Milburn:2000:az,Porras:2004:tt,Lin:2011:aa,Korenblit:2012:zz,Jurcevic:2014:uu,Hayes:2014:kk,Smith2016:im,Monroe:2019:za}. We will now briefly describe the relevant experimental procedures (see, e.g., Ref.~\cite{Monroe:2019:za} for details).

A common approach is to apply site-dependent optical Raman beams to selected ions, with beatnotes between the beams tuned to specified frequencies such as to give rise to the desired Hamiltonians \cite{Lee:2005:kk,Soderberg:2010:za,Bruzewicz:2019:aa,Monroe:2019:za}. When the beatnote for Raman beams focused on a given ion is tuned to the resonant frequency $\omega_0$ between the qubit levels, a Hamiltonian of the form
\begin{equation}\label{eq:3}
\op{H}_\phi = \frac{1}{2} \hbar \Omega \op{\sigma}_\phi
\end{equation}
can be realized \cite{Haljan:2005:km,Soderberg:2010:za,Monroe:2019:za}. Here $\Omega$ denotes the resonant Rabi frequency (which can be adjusted  by varying the detuning of the laser beams from the intermediary excited state connecting the qubit states via the Raman process \cite{Soderberg:2010:za,Monroe:2019:za}), and
\begin{equation}\label{eq:2}
\op{\sigma}_\phi =  \op{\sigma}_x \cos\phi - \op{\sigma}_y \sin\phi,
\end{equation}
where the angle $\phi$ can be precisely controlled via the phase of the Raman beatnote. If the beatnote is tuned away from resonance, one can make use of a differential ac Stark shift between the qubit levels to realize a Hamiltonian proportional to $\op{\sigma}_z$ \cite{Haffner:2003:uu,Leibfried:2003:mm,Lee:2016:in,Monroe:2019:za}. By addressing each ion with laser beams of specific intensity and detuning, a wide range of site-specific single-ion Hamiltonians can be implemented \cite{Lee:2016:in,Smith2016:im,Brydges:2019:aa,Maier:2019:uu,Monroe:2019:za}. For example, in Refs.~\cite{Smith2016:im,Brydges:2019:aa,Maier:2019:uu} this is achieved by deflecting a detuned laser beam from an acousto-optical deflector driven by a set of radio frequencies, generating independent, precisely controllable ac Stark shifts for each ion. This method for implementing tunable single-qubit Hamiltonians in ion traps has been used in several experiments to simulate disorder-inducing, site-specific transverse (axial) magnetic fields in many-spin systems \cite{Smith2016:im,Maier:2019:uu,Brydges:2019:aa,Monroe:2019:za}.  

Two-qubit interaction Hamiltonians of the form $\op{H} \propto (\bopvecgr{\sigma} \cdot \buvec{m}) \otimes (\bopvecgr{\sigma} \cdot \buvec{n})$, as needed for implementation of the system--probe Hamiltonian $\op{H}_m$ in Eq.~\eqref{eq:vdkkjl7}, can be realized in the same way as two-qubit gates in ion-trap quantum computation \cite{Sorensen:1999:za,Milburn:2000:az,Lee:2005:kk,Blatt:2008:uu,Soderberg:2010:za,Bruzewicz:2019:aa}. Just like the single-ion Hamiltonians, they are implemented via appropriately tuned laser beams applied to the ions \cite{Sorensen:1999:za,Milburn:2000:az,Lee:2005:kk,Blatt:2008:uu,Soderberg:2010:za,Bruzewicz:2019:aa,Monroe:2019:za}. Specifically, tuning the Raman beatnote to the vicinity of the phonon modes (the ``resonant regime'' \cite{Sorensen:1999:za,Milburn:2000:az}) gives rise to an effective spin--spin interaction between the ions mediated by the collective motional state of the ions \cite{Sorensen:1999:za,Milburn:2000:az,Lee:2005:kk,Blatt:2008:uu,Soderberg:2010:za,Bruzewicz:2019:aa}. For example, by simultaneously addressing two ions with bichromatic beatnotes symmetrically detuned from the blue and red vibrational sidebands, we can generate interaction Hamiltonians of the form \cite{Sorensen:1999:za,Lee:2005:kk,Blatt:2008:uu,Soderberg:2010:za,Monroe:2019:za}
\begin{equation}
\op{H} = \hbar J_{0} \op{\sigma}^{(1)}_\theta \op{\sigma}^{(2)}_\theta,
\end{equation}
where 
\begin{equation}\label{eq:4}
\op{\sigma}^{(i)}_\theta = \op{\sigma}^{(i)}_x\sin\theta+\op{\sigma}^{(i)}_y\cos\theta.
\end{equation}
The coupling strength $J_{0}$ can be precisely controlled via the detuning of the Raman beams \cite{Lee:2005:kk,Lin:2011:aa,Korenblit:2012:zz,Monroe:2019:za} or via local spatial control \cite{Hayes:2014:kk}, and the Bloch angle $\theta$ can be controlled by means of the phases of the beatnotes \cite{Monroe:2019:za}. Such interactions, applied to larger systems of trapped ions with each ion addressed locally by specific beatnotes and laser intensities, are also used in simulations of many-spin systems for implementing Ising-type Hamiltonians  of the form $\op{H} = \hbar \sum_{ij} J_{ij} \op{\sigma}^{(i)}_\theta \op{\sigma}^{(j)}_\theta$ \cite{Porras:2004:tt,Lin:2011:aa,Korenblit:2012:zz,Jurcevic:2014:uu,Smith2016:im,Brydges:2019:aa,Maier:2019:uu,Monroe:2019:za}. While in our application to protective measurement only two-qubit interactions are needed, such trapped-ion simulators are a particularly attractive platform for the experimental realization of the protective-measurement interaction due to their ability to precisely program and gate the desired Hamiltonians, with full control over the structure and strength of the interactions \cite{Lin:2011:aa,Korenblit:2012:zz,Hayes:2014:kk,Smith2016:im,Brydges:2019:aa,Maier:2019:uu,Monroe:2019:za}.

Thus, by addressing trapped ions with a combination of the laser fields we have described, we can simultaneously apply to the system qubit a local protection Hamiltonian proportional to $\op{\sigma}_\phi$ [see Eq.~\eqref{eq:2}] or $\op{\sigma}_z$ with adjustable strength, and to the system and probe qubits a variety of interaction Hamiltonians $\op{H}_m = \hbar J_0 (\bopvecgr{\sigma} \cdot \buvec{m}) \otimes (\bopvecgr{\sigma} \cdot \buvec{m})$ with adjustable strength $J_0$, representing protective measurements of different system observables $\bopvecgr{\sigma} \cdot \buvec{m}$. As an example, suppose we choose the beatnote phase such that $\phi=0$ in Eq.~\eqref{eq:3} and therefore the protection Hamiltonian becomes $\op{H}_S = \frac{1}{2} \hbar \Omega \op{\sigma}_x$. Then a continuous range of measurement orientations $\buvec{m}$ can be realized by choosing different values for the angle $\theta$ in Eq.~\eqref{eq:4}, which, as mentioned, can be done by tuning the beatnote phases for the Raman beams producing the two-qubit interaction. Note that $\frac{\pi}{2}-\theta$ is precisely the angle $\gamma$ defined in Sec.~\ref{sec:hamiltonian}, since $\gamma$ specifies the direction of the measured observable relative to the protection direction (here $\buvec{x}$). For an  interaction Hamiltonian of the form $(\bopvecgr{\sigma} \cdot \buvec{m}) \otimes (\bopvecgr{\sigma} \cdot \buvec{m})$, the rotation of the probe qubit will be around the same axis $\buvec{m}$ that defines the measurement. This, however, implies no loss of generality since the choice of the probe rotation axis is arbitrary and irrelevant to the physics of a protective measurement. 

Since one can address each ion individually without appreciably affecting other ions in the trap \cite{Smith2016:im,Bruzewicz:2019:aa,Monroe:2019:za}, we can realize $\op{H}_P\approx 0$ for the probe qubit as required for an optimal protective measurement [compare Eq.~\eqref{eq:vdkkjl7}]. Thus, we can avoid the complications arising from intrinsic probe dynamics as discussed in Sec.~\ref{sec:infl-intr-probe}. But this same site-specific addressing of the ions also enables us to experimentally explore, in a controlled way, the influence of intrinsic probe dynamics on a protective measurement, using the same techniques just described for the system qubit. By varying the parameter $\phi$ in Eq.~\eqref{eq:2} through adjustment of the Raman beatnote phase, we can experimentally test the influence of the probe dynamics not only for different strengths of the probe Hamiltonian, but also for different probe rotation axes as studied in Sec.~\ref{sec:infl-intr-probe}.

Experimentally available parameter values are well suited to the implementation of a protective measurement. As an example, let us consider the experiment described in Ref.~\cite{Brydges:2019:aa}. It uses a string of $^{40}$Ca$^+$ ions confined in a linear Paul trap, with the qubit states $\ket{0}$ and $\ket{1}$ represented by the Zeeman sublevels $\ket{S_{1/2},m_j=1/2}$ and $\ket{D_{5/2},m_j=5/2}$. The experiment uses a bichromatic laser beam to realize (here we consider only two neighboring ions in the trap) an interaction Hamiltonian $\op{H}_\text{int} = \hbar J_0 \op{\sigma}_x \op{\sigma}_x$ with $J_0\approx\unit[400]{s^{-1}}$. The site-specific ac Stark shift, implemented with a detuned laser beam deflected from an acousto-optical deflector, realizes a Hamiltonian $\op{H} = \hbar\Delta_i \op{\sigma}^{(i)}_z$ for each ion $i$. The strength $\Delta_i$ can be adjusted independently for each ion over the range $\Delta_i\in [0,6J_0]$. For application to an ideal protective measurement, we let $\Delta_1$ refer to the system qubit and $\Delta_2=0$ to the probe qubit (where we may also choose $\Delta_2\not= 0$ to implement intrinsic probe dynamics). The adjustable ratio $J_0/\Delta_1$ corresponds precisely to the measurement strength $\xi$ defined and used in Secs.~\ref{sec:model} and \ref{sec:nonid-meas}, and thus the experiment achieves measurement strengths as low as $\xi_\text{min}=0.17$. (Similarly, the experiment in Ref.~\cite{Smith2016:im}, which employs two hyperfine ``clock'' states of a $^{171}$Yb$^+$ ion as qubit levels, uses $\Delta_i\in [0,8J_0]$, giving $\xi_\text{min}=0.13$.) Thus, experiments of this kind not only allow one to adjust the measurement strength, but they also reach the desired weak-measurement regime with $\xi$ substantially below 1. The value $J_0 \approx \unit[400]{s^{-1}}$ in the experiment of Ref.~\cite{Brydges:2019:aa} implies a timescale $T$ for the interaction on the order of a few milliseconds, which can be precisely controlled and resolved.

The influence of an environment on the protective-measurement process as discussed in Sec.~\ref{sec:infl-inter-with} can also be experimentally investigated with such experiments. For example, in Ref.~\cite{Maier:2019:uu} tunable dephasing between the qubit states was experimentally introduced through controlled temporal modulations of the ac Stark shifts for each ion. This amounts to adding a stationary noise term $\hbar W_i(t) \op{\sigma}_z$ with adjustable spectral power to the self-Hamiltonian for each ion.

\section{\label{sec:discussion}Discussion}

We have considered a variant of a protective qubit measurement in which the probe is represented by a two-state system, rather than by a continuous phase-space degree of freedom. This model reproduces the essence of a protective measurement, namely, the transfer of information about the expectation value for an observable of the system while the state of the system is left approximately undisturbed. One motivation for considering a qubit probe is the relative ease with which the protective measurement may be experimentally implemented.

The evolution of the system is seen as a precession of the Bloch vector around the axis defined by the protection field, while the Bloch vector for the probe is gradually rotated around an arbitrarily chosen axis, with the rotation angle encoding the desired expectation value for the system. Analysis of these dynamics demonstrates that even for an only moderately weak measurement, we can achieve both small state disturbance for the system and a faithful pointer shift for the probe. Our results also show that in cases where the intrinsic dynamics of the probe during the measurement are significant, the measurement interaction should be chosen such that the probe rotation is around the same axis as the rotation produced by the intrinsic dynamics, since in this case the effect of the intrinsic evolution on the measurement outcome may be fully compensated for by an appropriately chosen counter-rotation. Furthermore, our analysis illustrates how the influence of interactions with an environment on the measurement depends on the form of the environmental monitoring process, and how the coupling of the system to the environment affects the evolution of the probe even when the latter is not directly interacting with an environment.  
 
Because the desired expectation value is encoded in the rotation angle of the probe qubit, readout of this information requires the measurement of an expectation value on the probe, and thus the accumulation of probe statistics from a series of system--probe interactions and measurements of the probe states. We have shown that even with such repeated measurements, low cumulative state disturbance and proper probe rotation can be maintained, provided the measurements are carried out in the weak regime (as is generally assumed for a protective measurement). 

We may also compare the need for measuring an expectation value on the probe to the situation encountered for a phase-space probe \cite{Dass:1999:az}. There, the change in the location of the center of the pointer wave packet in the relevant variable (position or momentum) represents the pointer shift. This location (or, alternatively, its change) is given by the expectation value for an appropriate pointer observable. Thus, in order to precisely resolve the pointer shift, one will need to measure an expectation value on the probe. This implies having to resort to measurements on an ensemble of probes, each of which has interacted with the system via the protective-measurement coupling, or otherwise perform multiple measurements on the same probe, possibly by using quantum nondemolition schemes (see Sec.~IV of Ref.~\cite{Dass:1999:az} for a discussion of these options). In this way, the situation with regard to the readout of a phase-space probe is similar to that for a qubit probe. In practice, the main difference is that if the pointer wave packet for the phase-space probe starts out sufficiently narrow and remains so throughout the measurement \footnote{See Ref.~\cite{Dass:1999:az} for an analysis of wave-packet spreading during a protective measurement.} then a single measurement on the pointer may provide an estimate of the wave-packet center with satisfactory precision (as determined by the width of the wave packet), simply because the measurement outcome is likely to fall in the vicinity of the wave-packet center. By contrast, for a qubit probe an ensemble of probe measurements is always required, since any single (projective) measurement of a qubit observable will give just one of two possible outcomes. Apart from this difference, a protective measurement with a qubit probe has the same essential features as that with a phase-space probe. In particular, it requires only a single system qubit, the disturbance of the system can be made arbitrarily small, and in each iteration of the probe preparation--interaction--readout cycle, the probe is deterministically brought to the same final state, meaning that complete information about the expectation value of the system is deterministically transferred to the pointer during each individual interaction with the probe. 

One may wonder about the advantage of the protective-measurement scheme given that one needs to measure an expectation value on the probe in order to infer an expectation value of the system. To recognize this advantage, it is important to remember that system and probe play fundamentally different roles. The system is in an unknown quantum state, which we would like to determine without appreciably changing it (we refer here to the task of measuring a quantum state rather than an expectation value, since measurement of expectation values allows reconstruction of the state). The probe, by contrast, merely plays the role of an ancilla. It can be repeatedly prepared in an arbitrary state and subjected to an arbitrary readout measurement, and the disturbance of its state by such measurements is irrelevant. So while one needs to repeatedly measure the ancilla to obtain an expectation value for it, what is achieved in this way is a measurement of the state of the system while disturbing it arbitrarily little, a nontrivial task \cite{Aharonov:1993:qa,Aharonov:1993:jm,Dass:1999:az,Vaidman:2009:po,Gao:2014:cu,Genovese:2017:zz,Qureshi:2015:jj}.

A distinct advantage of using a qubit probe is the amenability of the resulting protective-measurement scheme to experimental implementation using current technology. Given that protective measurements (other than the conceptually quite distinct quantum Zeno version \cite{Piacentini:2017:oo}) have so far eluded experimental realization, this is an important asset. As we have discussed, existing ion-trap experiments already offer all the techniques and tools needed for an experimental implementation of a protective measurement using the scheme described here, including high-fidelity state preparation and readout, engineering of the relevant site-specific Hamiltonians with precise control over their structure and strength, and parameters well within the regime suitable for protective measurements. Ion-trap quantum simulators of many-spin systems \cite{Porras:2004:tt,Lin:2011:aa,Korenblit:2012:zz,Hayes:2014:kk,Smith2016:im,Monroe:2019:za} provide an especially promising experimental platform, since they allow one to precisely design and tune the Hamiltonians at the level of individual ions. In this way, it should be possible not only to realize a single protective measurement, but also to experimentally explore such measurements in quantitative detail along the lines of the analysis given in this paper. By varying the parameters of the optical fields interacting with the trapped ions, one may implement protective measurements of different observables, trace the transition from ideal (state-preserving) to nonideal protective measurements, study and control the influence of intrinsic probe dynamics, and investigate the effect of environmental interactions on the measurement.


\begin{thebibliography}{48}%
\makeatletter
\providecommand \@ifxundefined [1]{%
 \@ifx{#1\undefined}
}%
\providecommand \@ifnum [1]{%
 \ifnum #1\expandafter \@firstoftwo
 \else \expandafter \@secondoftwo
 \fi
}%
\providecommand \@ifx [1]{%
 \ifx #1\expandafter \@firstoftwo
 \else \expandafter \@secondoftwo
 \fi
}%
\providecommand \natexlab [1]{#1}%
\providecommand \enquote  [1]{``#1''}%
\providecommand \bibnamefont  [1]{#1}%
\providecommand \bibfnamefont [1]{#1}%
\providecommand \citenamefont [1]{#1}%
\providecommand \href@noop [0]{\@secondoftwo}%
\providecommand \href [0]{\begingroup \@sanitize@url \@href}%
\providecommand \@href[1]{\@@startlink{#1}\@@href}%
\providecommand \@@href[1]{\endgroup#1\@@endlink}%
\providecommand \@sanitize@url [0]{\catcode `\\12\catcode `\$12\catcode
  `\&12\catcode `\#12\catcode `\^12\catcode `\_12\catcode `\%12\relax}%
\providecommand \@@startlink[1]{}%
\providecommand \@@endlink[0]{}%
\providecommand \url  [0]{\begingroup\@sanitize@url \@url }%
\providecommand \@url [1]{\endgroup\@href {#1}{\urlprefix }}%
\providecommand \urlprefix  [0]{URL }%
\providecommand \Eprint [0]{\href }%
\providecommand \doibase [0]{http://dx.doi.org/}%
\providecommand \selectlanguage [0]{\@gobble}%
\providecommand \bibinfo  [0]{\@secondoftwo}%
\providecommand \bibfield  [0]{\@secondoftwo}%
\providecommand \translation [1]{[#1]}%
\providecommand \BibitemOpen [0]{}%
\providecommand \bibitemStop [0]{}%
\providecommand \bibitemNoStop [0]{.\EOS\space}%
\providecommand \EOS [0]{\spacefactor3000\relax}%
\providecommand \BibitemShut  [1]{\csname bibitem#1\endcsname}%
\let\auto@bib@innerbib\@empty
\bibitem [{\citenamefont {Dressel}\ \emph {et~al.}(2014)\citenamefont
  {Dressel}, \citenamefont {Malik}, \citenamefont {Miatto}, \citenamefont
  {Jordan},\ and\ \citenamefont {Boyd}}]{Dressel:2014:uu}%
  \BibitemOpen
  \bibfield  {author} {\bibinfo {author} {\bibfnamefont {J.}~\bibnamefont
  {Dressel}}, \bibinfo {author} {\bibfnamefont {M.}~\bibnamefont {Malik}},
  \bibinfo {author} {\bibfnamefont {F.~M.}\ \bibnamefont {Miatto}}, \bibinfo
  {author} {\bibfnamefont {A.~N.}\ \bibnamefont {Jordan}}, \ and\ \bibinfo
  {author} {\bibfnamefont {R.~W.}\ \bibnamefont {Boyd}},\ }\href {\doibase
  10.1103/RevModPhys.86.307} {\bibfield  {journal} {\bibinfo  {journal} {Rev.
  Mod. Phys.}\ }\textbf {\bibinfo {volume} {86}},\ \bibinfo {pages} {307}
  (\bibinfo {year} {2014})}\BibitemShut {NoStop}%
\bibitem [{\citenamefont {Gao}(2014)}]{Gao:2014:cu}%
  \BibitemOpen
  \bibinfo {editor} {\bibfnamefont {S.}~\bibnamefont {Gao}},\ ed.,\ \href@noop
  {} {\emph {\bibinfo {title} {Protective Measurement and Quantum Reality:
  Towards a New Understanding of Quantum Mechanics}}}\ (\bibinfo  {publisher}
  {Cambridge University Press},\ \bibinfo {address} {Cambridge},\ \bibinfo
  {year} {2014})\BibitemShut {NoStop}%
\bibitem [{\citenamefont {Aharonov}\ and\ \citenamefont
  {Vaidman}(1993)}]{Aharonov:1993:qa}%
  \BibitemOpen
  \bibfield  {author} {\bibinfo {author} {\bibfnamefont {Y.}~\bibnamefont
  {Aharonov}}\ and\ \bibinfo {author} {\bibfnamefont {L.}~\bibnamefont
  {Vaidman}},\ }\href {\doibase 10.1016/0375-9601(93)90724-E} {\bibfield
  {journal} {\bibinfo  {journal} {Phys. Lett. A}\ }\textbf {\bibinfo {volume}
  {178}},\ \bibinfo {pages} {38} (\bibinfo {year} {1993})}\BibitemShut
  {NoStop}%
\bibitem [{\citenamefont {Aharonov}\ \emph {et~al.}(1993)\citenamefont
  {Aharonov}, \citenamefont {Anandan},\ and\ \citenamefont
  {Vaidman}}]{Aharonov:1993:jm}%
  \BibitemOpen
  \bibfield  {author} {\bibinfo {author} {\bibfnamefont {Y.}~\bibnamefont
  {Aharonov}}, \bibinfo {author} {\bibfnamefont {J.}~\bibnamefont {Anandan}}, \
  and\ \bibinfo {author} {\bibfnamefont {L.}~\bibnamefont {Vaidman}},\ }\href
  {\doibase 10.1103/PhysRevA.47.4616} {\bibfield  {journal} {\bibinfo
  {journal} {Phys. Rev. A}\ }\textbf {\bibinfo {volume} {47}},\ \bibinfo
  {pages} {4616} (\bibinfo {year} {1993})}\BibitemShut {NoStop}%
\bibitem [{\citenamefont {{Hari Dass}}\ and\ \citenamefont
  {Qureshi}(1999)}]{Dass:1999:az}%
  \BibitemOpen
  \bibfield  {author} {\bibinfo {author} {\bibfnamefont {N.~D.}\ \bibnamefont
  {{Hari Dass}}}\ and\ \bibinfo {author} {\bibfnamefont {T.}~\bibnamefont
  {Qureshi}},\ }\href {\doibase 10.1103/PhysRevA.59.2590} {\bibfield  {journal}
  {\bibinfo  {journal} {Phys. Rev. A}\ }\textbf {\bibinfo {volume} {59}},\
  \bibinfo {pages} {2590} (\bibinfo {year} {1999})}\BibitemShut {NoStop}%
\bibitem [{\citenamefont {Vaidman}(2009)}]{Vaidman:2009:po}%
  \BibitemOpen
  \bibfield  {author} {\bibinfo {author} {\bibfnamefont {L.}~\bibnamefont
  {Vaidman}},\ }in\ \href@noop {} {\emph {\bibinfo {booktitle} {Compendium of
  Quantum Physics: Concepts, Experiments, History and Philosophy}}},\ \bibinfo
  {editor} {edited by\ \bibinfo {editor} {\bibfnamefont {D.}~\bibnamefont
  {Greenberger}}, \bibinfo {editor} {\bibfnamefont {K.}~\bibnamefont
  {Hentschel}}, \ and\ \bibinfo {editor} {\bibfnamefont {F.}~\bibnamefont
  {Weinert}}}\ (\bibinfo  {publisher} {Springer},\ \bibinfo {address}
  {Berlin/Heidelberg},\ \bibinfo {year} {2009}), pp.\ \bibinfo {pages}
  {505--508}\BibitemShut {NoStop}%
\bibitem [{\citenamefont {Genovese}(2017)}]{Genovese:2017:zz}%
  \BibitemOpen
  \bibfield  {author} {\bibinfo {author} {\bibfnamefont {M.}~\bibnamefont
  {Genovese}},\ }\href {\doibase 10.1088/1742-6596/880/1/012012} {\bibfield
  {journal} {\bibinfo  {journal} {J. Phys.: Conf. Ser.}\ }\textbf {\bibinfo
  {volume} {880}},\ \bibinfo {pages} {012012} (\bibinfo {year}
  {2017})}\BibitemShut {NoStop}%
\bibitem [{\citenamefont {Qureshi}\ and\ \citenamefont {{Hari
  Dass}}(2015)}]{Qureshi:2015:jj}%
  \BibitemOpen
  \bibfield  {author} {\bibinfo {author} {\bibfnamefont {T.}~\bibnamefont
  {Qureshi}}\ and\ \bibinfo {author} {\bibfnamefont {N.~D.}\ \bibnamefont
  {{Hari Dass}}},\ }\href {\doibase 10.18520/v109/i11/2023-2028} {\bibfield
  {journal} {\bibinfo  {journal} {Curr. Sci.}\ }\textbf {\bibinfo {volume}
  {109}},\ \bibinfo {pages} {2023} (\bibinfo {year} {2015})}\BibitemShut
  {NoStop}%
\bibitem [{\citenamefont {Anandan}\ and\ \citenamefont
  {Vaidman}(1996)}]{Aharonov:1996:fp}%
  \BibitemOpen
  \bibfield  {author} {\bibinfo {author} {\bibfnamefont {Y.~A.~J.}\
  \bibnamefont {Anandan}}\ and\ \bibinfo {author} {\bibfnamefont
  {L.}~\bibnamefont {Vaidman}},\ }\href {\doibase 10.1007/BF02058891}
  {\bibfield  {journal} {\bibinfo  {journal} {Found. Phys.}\ }\textbf {\bibinfo
  {volume} {26}},\ \bibinfo {pages} {117} (\bibinfo {year} {1996})}\BibitemShut
  {NoStop}%
\bibitem [{\citenamefont {Auletta}(2014)}]{Auletta:2014:yy}%
  \BibitemOpen
  \bibfield  {author} {\bibinfo {author} {\bibfnamefont {G.}~\bibnamefont
  {Auletta}},\ }in\ Ref.~\cite{Gao:2014:cu}, pp.\
  \bibinfo {pages} {39--62}\BibitemShut {NoStop}%
\bibitem [{\citenamefont {Di{\'o}si}(2014)}]{Diosi:2014:yy}%
  \BibitemOpen
  \bibfield  {author} {\bibinfo {author} {\bibfnamefont {L.}~\bibnamefont
  {Di{\'o}si}},\ }in\ Ref.~\cite{Gao:2014:cu}, pp.\
  \bibinfo {pages} {63--67}\BibitemShut {NoStop}%
\bibitem [{\citenamefont {Aharonov}\ and\ \citenamefont
  {Cohen}(2014)}]{Aharonov:2014:yy}%
  \BibitemOpen
  \bibfield  {author} {\bibinfo {author} {\bibfnamefont {Y.}~\bibnamefont
  {Aharonov}}\ and\ \bibinfo {author} {\bibfnamefont {E.}~\bibnamefont
  {Cohen}},\ }in\ Ref.~\cite{Gao:2014:cu}, pp.\
  \bibinfo {pages} {28--38}\BibitemShut {NoStop}%
\bibitem [{\citenamefont {Schlosshauer}(2016)}]{Schlosshauer:2016:uu}%
  \BibitemOpen
  \bibfield  {author} {\bibinfo {author} {\bibfnamefont {M.}~\bibnamefont
  {Schlosshauer}},\ }\href {\doibase 10.1103/PhysRevA.93.012115} {\bibfield
  {journal} {\bibinfo  {journal} {Phys. Rev. A}\ }\textbf {\bibinfo {volume}
  {93}},\ \bibinfo {pages} {012115} (\bibinfo {year} {2016})}\BibitemShut
  {NoStop}%
\bibitem [{\citenamefont {Alter}\ and\ \citenamefont
  {Yamamoto}(1997)}]{Alter:1997:oo}%
  \BibitemOpen
  \bibfield  {author} {\bibinfo {author} {\bibfnamefont {O.}~\bibnamefont
  {Alter}}\ and\ \bibinfo {author} {\bibfnamefont {Y.}~\bibnamefont
  {Yamamoto}},\ }\href {\doibase 10.1103/PhysRevA.56.1057} {\bibfield
  {journal} {\bibinfo  {journal} {Phys. Rev. A}\ }\textbf {\bibinfo {volume}
  {56}},\ \bibinfo {pages} {1057} (\bibinfo {year} {1997})}\BibitemShut
  {NoStop}%
\bibitem [{\citenamefont {Aharonov}\ and\ \citenamefont
  {Vaidman}(1996)}]{Aharonov:1996:ii}%
  \BibitemOpen
  \bibfield  {author} {\bibinfo {author} {\bibfnamefont {Y.}~\bibnamefont
  {Aharonov}}\ and\ \bibinfo {author} {\bibfnamefont {L.}~\bibnamefont
  {Vaidman}},\ }in\ \href@noop {} {\emph {\bibinfo {booktitle} {Bohmian
  Mechanics and Quantum Theory: An Appraisal}}},\ \bibinfo {editor} {edited by\
  \bibinfo {editor} {\bibfnamefont {J.~T.}\ \bibnamefont {Cushing}}, \bibinfo
  {editor} {\bibfnamefont {A.}~\bibnamefont {Fine}}, \ and\ \bibinfo {editor}
  {\bibfnamefont {S.}~\bibnamefont {Goldstein}}}\ (\bibinfo  {publisher}
  {Kluwer},\ \bibinfo {address} {Dordrecht},\ \bibinfo {year} {1996}), pp.\
  \bibinfo {pages} {141--154}\BibitemShut {NoStop}%
\bibitem [{\citenamefont {Aharonov}\ \emph {et~al.}(1999)\citenamefont
  {Aharonov}, \citenamefont {Englert},\ and\ \citenamefont
  {Scully}}]{Aharonov:1999:uu}%
  \BibitemOpen
  \bibfield  {author} {\bibinfo {author} {\bibfnamefont {Y.}~\bibnamefont
  {Aharonov}}, \bibinfo {author} {\bibfnamefont {B.~G.}\ \bibnamefont
  {Englert}}, \ and\ \bibinfo {author} {\bibfnamefont {M.~O.}\ \bibnamefont
  {Scully}},\ }\href {\doibase 10.1016/S0375-9601(99)00628-3} {\bibfield
  {journal} {\bibinfo  {journal} {Phys. Lett. A}\ }\textbf {\bibinfo {volume}
  {263}},\ \bibinfo {pages} {137} (\bibinfo {year} {1999})}\BibitemShut
  {NoStop}%
\bibitem [{\citenamefont {Schlosshauer}(2015)}]{Schlosshauer:2015:uu}%
  \BibitemOpen
  \bibfield  {author} {\bibinfo {author} {\bibfnamefont {M.}~\bibnamefont
  {Schlosshauer}},\ }\href {\doibase 10.1103/PhysRevA.92.062116} {\bibfield
  {journal} {\bibinfo  {journal} {Phys. Rev. A}\ }\textbf {\bibinfo {volume}
  {92}},\ \bibinfo {pages} {062116} (\bibinfo {year} {2015})}\BibitemShut
  {NoStop}%
\bibitem [{\citenamefont {{Hari Dass}}(1999)}]{Dass:1999:le}%
  \BibitemOpen
  \bibfield  {author} {\bibinfo {author} {\bibfnamefont {N.~D.}\ \bibnamefont
  {{Hari Dass}}},\ }\href@noop {} \Eprint
  {http://arxiv.org/abs/arXiv:quant-ph/9908085} {arXiv:quant-ph/9908085}
  \BibitemShut {NoStop}%
\bibitem [{\citenamefont {Piacentini}\ \emph {et~al.}(2017)\citenamefont
  {Piacentini}, \citenamefont {Avella}, \citenamefont {Rebufello},
  \citenamefont {Lussana}, \citenamefont {Villa}, \citenamefont {Tosi},
  \citenamefont {Gramegna}, \citenamefont {Brida}, \citenamefont {Cohen},
  \citenamefont {Vaidman}, \citenamefont {Degiovanni},\ and\ \citenamefont
  {Genovese}}]{Piacentini:2017:oo}%
  \BibitemOpen
  \bibfield  {author} {\bibinfo {author} {\bibfnamefont {F.}~\bibnamefont
  {Piacentini}}, \bibinfo {author} {\bibfnamefont {A.}~\bibnamefont {Avella}},
  \bibinfo {author} {\bibfnamefont {E.}~\bibnamefont {Rebufello}}, \bibinfo
  {author} {\bibfnamefont {R.}~\bibnamefont {Lussana}}, \bibinfo {author}
  {\bibfnamefont {F.}~\bibnamefont {Villa}}, \bibinfo {author} {\bibfnamefont
  {A.}~\bibnamefont {Tosi}}, \bibinfo {author} {\bibfnamefont {M.}~\bibnamefont
  {Gramegna}}, \bibinfo {author} {\bibfnamefont {G.}~\bibnamefont {Brida}},
  \bibinfo {author} {\bibfnamefont {E.}~\bibnamefont {Cohen}}, \bibinfo
  {author} {\bibfnamefont {L.}~\bibnamefont {Vaidman}}, \bibinfo {author}
  {\bibfnamefont {I.~P.}\ \bibnamefont {Degiovanni}}, \ and\ \bibinfo {author}
  {\bibfnamefont {M.}~\bibnamefont {Genovese}},\ }\href {\doibase
  10.1038/nphys4223} {\bibfield  {journal} {\bibinfo  {journal} {Nature Phys.}\
  }\textbf {\bibinfo {volume} {13}},\ \bibinfo {pages} {1191} (\bibinfo {year}
  {2017})}\BibitemShut {NoStop}%
\bibitem [{Note1()}]{Note1}%
  \BibitemOpen
  \bibinfo {note} {In the standard scheme we use in this paper, Alice can give a protected state
  (that she may know) to Bob, who can then apply the measurement interaction to
  obtain expectation values without needing to know the protection procedure or
  the protected state. In this sense, Bob can measure an unknown quantum state.
  By contrast, in the quantum Zeno version, Bob needs to apply projections on
  the state throughout the measurement interaction, and would thus need to know
  the state all along.}\BibitemShut {Stop}%
\bibitem [{\citenamefont {Schlosshauer}(2014)}]{Schlosshauer:2014:pm}%
  \BibitemOpen
  \bibfield  {author} {\bibinfo {author} {\bibfnamefont {M.}~\bibnamefont
  {Schlosshauer}},\ }\href {\doibase 10.1103/PhysRevA.90.052106} {\bibfield
  {journal} {\bibinfo  {journal} {Phys. Rev. A}\ }\textbf {\bibinfo {volume}
  {90}},\ \bibinfo {pages} {052106} (\bibinfo {year} {2014})}\BibitemShut
  {NoStop}%
\bibitem [{\citenamefont {Bruzewicz}\ \emph {et~al.}(2019)\citenamefont
  {Bruzewicz}, \citenamefont {Chiaverini}, \citenamefont {McConnell},\ and\
  \citenamefont {Sage}}]{Bruzewicz:2019:aa}%
  \BibitemOpen
  \bibfield  {author} {\bibinfo {author} {\bibfnamefont {C.~D.}\ \bibnamefont
  {Bruzewicz}}, \bibinfo {author} {\bibfnamefont {J.}~\bibnamefont
  {Chiaverini}}, \bibinfo {author} {\bibfnamefont {R.}~\bibnamefont
  {McConnell}}, \ and\ \bibinfo {author} {\bibfnamefont {J.~M.}\ \bibnamefont
  {Sage}},\ }\href {\doibase 10.1063/1.5088164} {\bibfield  {journal} {\bibinfo
   {journal} {Appl. Phys. Rev.}\ }\textbf {\bibinfo {volume} {6}},\ \bibinfo
  {pages} {021314} (\bibinfo {year} {2019})}\BibitemShut {NoStop}%
\bibitem [{\citenamefont {Milburn}\ \emph {et~al.}(2000)\citenamefont
  {Milburn}, \citenamefont {Schneider},\ and\ \citenamefont
  {James}}]{Milburn:2000:az}%
  \BibitemOpen
  \bibfield  {author} {\bibinfo {author} {\bibfnamefont {G.~J.}\ \bibnamefont
  {Milburn}}, \bibinfo {author} {\bibfnamefont {S.}~\bibnamefont {Schneider}},
  \ and\ \bibinfo {author} {\bibfnamefont {D.~F.~V.}\ \bibnamefont {James}},\
  }\href@noop {} {\bibfield  {journal} {\bibinfo  {journal} {Fortschr. Phys.}\
  }\textbf {\bibinfo {volume} {48}},\ \bibinfo {pages} {9} (\bibinfo {year}
  {2000})}\BibitemShut {NoStop}%
\bibitem [{\citenamefont {Porras}\ and\ \citenamefont
  {Cirac}(2004)}]{Porras:2004:tt}%
  \BibitemOpen
  \bibfield  {author} {\bibinfo {author} {\bibfnamefont {D.}~\bibnamefont
  {Porras}}\ and\ \bibinfo {author} {\bibfnamefont {J.~I.}\ \bibnamefont
  {Cirac}},\ }\href {\doibase 10.1103/PhysRevLett.92.207901} {\bibfield
  {journal} {\bibinfo  {journal} {Phys. Rev. Lett.}\ }\textbf {\bibinfo
  {volume} {92}},\ \bibinfo {pages} {207901} (\bibinfo {year}
  {2004})}\BibitemShut {NoStop}%
\bibitem [{\citenamefont {Lin}\ \emph {et~al.}(2011)\citenamefont {Lin},
  \citenamefont {Monroe},\ and\ \citenamefont {Duan}}]{Lin:2011:aa}%
  \BibitemOpen
  \bibfield  {author} {\bibinfo {author} {\bibfnamefont {G.-D.}\ \bibnamefont
  {Lin}}, \bibinfo {author} {\bibfnamefont {C.}~\bibnamefont {Monroe}}, \ and\
  \bibinfo {author} {\bibfnamefont {L.-M.}\ \bibnamefont {Duan}},\ }\href
  {\doibase 10.1103/PhysRevLett.106.230402} {\bibfield  {journal} {\bibinfo
  {journal} {Phys. Rev. Lett.}\ }\textbf {\bibinfo {volume} {106}},\ \bibinfo
  {pages} {230402} (\bibinfo {year} {2011})}\BibitemShut {NoStop}%
\bibitem [{\citenamefont {Korenblit}\ \emph {et~al.}(2012)\citenamefont
  {Korenblit}, \citenamefont {Kafri}, \citenamefont {Campbell}, \citenamefont
  {Islam}, \citenamefont {Edwards}, \citenamefont {Gong}, \citenamefont {Lin},
  \citenamefont {Duan}, \citenamefont {Kim}, \citenamefont {Kim},\ and\
  \citenamefont {Monroe}}]{Korenblit:2012:zz}%
  \BibitemOpen
  \bibfield  {author} {\bibinfo {author} {\bibfnamefont {S.}~\bibnamefont
  {Korenblit}}, \bibinfo {author} {\bibfnamefont {D.}~\bibnamefont {Kafri}},
  \bibinfo {author} {\bibfnamefont {W.~C.}\ \bibnamefont {Campbell}}, \bibinfo
  {author} {\bibfnamefont {R.}~\bibnamefont {Islam}}, \bibinfo {author}
  {\bibfnamefont {E.~E.}\ \bibnamefont {Edwards}}, \bibinfo {author}
  {\bibfnamefont {Z.-X.}\ \bibnamefont {Gong}}, \bibinfo {author}
  {\bibfnamefont {G.-D.}\ \bibnamefont {Lin}}, \bibinfo {author} {\bibfnamefont
  {L.-M.}\ \bibnamefont {Duan}}, \bibinfo {author} {\bibfnamefont
  {J.}~\bibnamefont {Kim}}, \bibinfo {author} {\bibfnamefont {K.}~\bibnamefont
  {Kim}}, \ and\ \bibinfo {author} {\bibfnamefont {C.}~\bibnamefont {Monroe}},\
  }\href {\doibase 10.1088/1367-2630/14/9/095024} {\bibfield  {journal}
  {\bibinfo  {journal} {New J. Phys.}\ }\textbf {\bibinfo {volume} {14}},\
  \bibinfo {pages} {095024} (\bibinfo {year} {2012})}\BibitemShut {NoStop}%
\bibitem [{\citenamefont {Jurcevic}\ \emph {et~al.}(2014)\citenamefont
  {Jurcevic}, \citenamefont {Lanyon}, \citenamefont {Hauke}, \citenamefont
  {Hempel}, \citenamefont {Zoller}, \citenamefont {Blatt},\ and\ \citenamefont
  {Roos}}]{Jurcevic:2014:uu}%
  \BibitemOpen
  \bibfield  {author} {\bibinfo {author} {\bibfnamefont {P.}~\bibnamefont
  {Jurcevic}}, \bibinfo {author} {\bibfnamefont {B.~P.}\ \bibnamefont
  {Lanyon}}, \bibinfo {author} {\bibfnamefont {P.}~\bibnamefont {Hauke}},
  \bibinfo {author} {\bibfnamefont {C.}~\bibnamefont {Hempel}}, \bibinfo
  {author} {\bibfnamefont {P.}~\bibnamefont {Zoller}}, \bibinfo {author}
  {\bibfnamefont {R.}~\bibnamefont {Blatt}}, \ and\ \bibinfo {author}
  {\bibfnamefont {C.~F.}\ \bibnamefont {Roos}},\ }\href {\doibase
  10.1038/nature13461} {\bibfield  {journal} {\bibinfo  {journal} {Nature}\
  }\textbf {\bibinfo {volume} {511}},\ \bibinfo {pages} {202} (\bibinfo {year}
  {2014})}\BibitemShut {NoStop}%
\bibitem [{\citenamefont {Hayes}\ \emph {et~al.}(2014)\citenamefont {Hayes},
  \citenamefont {Flammia},\ and\ \citenamefont {Biercuk}}]{Hayes:2014:kk}%
  \BibitemOpen
  \bibfield  {author} {\bibinfo {author} {\bibfnamefont {D.}~\bibnamefont
  {Hayes}}, \bibinfo {author} {\bibfnamefont {S.~T.}\ \bibnamefont {Flammia}},
  \ and\ \bibinfo {author} {\bibfnamefont {M.~J.}\ \bibnamefont {Biercuk}},\
  }\href {\doibase 10.1088/1367-2630/16/8/083027} {\bibfield  {journal}
  {\bibinfo  {journal} {New J. Phys.}\ }\textbf {\bibinfo {volume} {16}},\
  \bibinfo {pages} {083027} (\bibinfo {year} {2014})}\BibitemShut {NoStop}%
\bibitem [{\citenamefont {Smith}\ \emph {et~al.}(2016)\citenamefont {Smith},
  \citenamefont {Lee}, \citenamefont {Richerme}, \citenamefont {Neyenhuis},
  \citenamefont {Hess}, \citenamefont {Hauke}, \citenamefont {Heyl},
  \citenamefont {Huse},\ and\ \citenamefont {Monroe}}]{Smith2016:im}%
  \BibitemOpen
  \bibfield  {author} {\bibinfo {author} {\bibfnamefont {J.}~\bibnamefont
  {Smith}}, \bibinfo {author} {\bibfnamefont {A.}~\bibnamefont {Lee}}, \bibinfo
  {author} {\bibfnamefont {P.}~\bibnamefont {Richerme}}, \bibinfo {author}
  {\bibfnamefont {B.}~\bibnamefont {Neyenhuis}}, \bibinfo {author}
  {\bibfnamefont {P.~W.}\ \bibnamefont {Hess}}, \bibinfo {author}
  {\bibfnamefont {P.}~\bibnamefont {Hauke}}, \bibinfo {author} {\bibfnamefont
  {M.}~\bibnamefont {Heyl}}, \bibinfo {author} {\bibfnamefont {D.~A.}\
  \bibnamefont {Huse}}, \ and\ \bibinfo {author} {\bibfnamefont
  {C.}~\bibnamefont {Monroe}},\ }\href {\doibase 10.1038/nphys3783} {\bibfield
  {journal} {\bibinfo  {journal} {Nature Phys.}\ }\textbf {\bibinfo {volume}
  {12}},\ \bibinfo {pages} {907} (\bibinfo {year} {2016})}\BibitemShut
  {NoStop}%
\bibitem [{\citenamefont {Monroe}\ \emph {et~al.}(2019)\citenamefont {Monroe},
  \citenamefont {Campbell}, \citenamefont {Duan}, \citenamefont {Gong},
  \citenamefont {Gorshkov}, \citenamefont {Hess}, \citenamefont {Islam},
  \citenamefont {Kim}, \citenamefont {Pagano}, \citenamefont {Richerme},
  \citenamefont {Senko},\ and\ \citenamefont {Yao}}]{Monroe:2019:za}%
  \BibitemOpen
  \bibfield  {author} {\bibinfo {author} {\bibfnamefont {C.}~\bibnamefont
  {Monroe}}, \bibinfo {author} {\bibfnamefont {W.~C.}\ \bibnamefont
  {Campbell}}, \bibinfo {author} {\bibfnamefont {L.~M.}\ \bibnamefont {Duan}},
  \bibinfo {author} {\bibfnamefont {Z.~X.}\ \bibnamefont {Gong}}, \bibinfo
  {author} {\bibfnamefont {A.~V.}\ \bibnamefont {Gorshkov}}, \bibinfo {author}
  {\bibfnamefont {P.}~\bibnamefont {Hess}}, \bibinfo {author} {\bibfnamefont
  {R.}~\bibnamefont {Islam}}, \bibinfo {author} {\bibfnamefont
  {K.}~\bibnamefont {Kim}}, \bibinfo {author} {\bibfnamefont {G.}~\bibnamefont
  {Pagano}}, \bibinfo {author} {\bibfnamefont {P.}~\bibnamefont {Richerme}},
  \bibinfo {author} {\bibfnamefont {C.}~\bibnamefont {Senko}}, \ and\ \bibinfo
  {author} {\bibfnamefont {N.~Y.}\ \bibnamefont {Yao}},}\href@noop {} {}
\Eprint {http://arxiv.org/abs/1912.07845}
  {arXiv:1912.07845 [quant-ph]} \BibitemShut {NoStop}%
\bibitem [{\citenamefont {Noek}\ \emph {et~al.}(2013)\citenamefont {Noek},
  \citenamefont {Vrijsen}, \citenamefont {Gaultney}, \citenamefont {Mount},
  \citenamefont {Kim}, \citenamefont {Maunz},\ and\ \citenamefont
  {Kim}}]{Noek:2013:uu}%
  \BibitemOpen
  \bibfield  {author} {\bibinfo {author} {\bibfnamefont {R.}~\bibnamefont
  {Noek}}, \bibinfo {author} {\bibfnamefont {G.}~\bibnamefont {Vrijsen}},
  \bibinfo {author} {\bibfnamefont {D.}~\bibnamefont {Gaultney}}, \bibinfo
  {author} {\bibfnamefont {E.}~\bibnamefont {Mount}}, \bibinfo {author}
  {\bibfnamefont {T.}~\bibnamefont {Kim}}, \bibinfo {author} {\bibfnamefont
  {P.}~\bibnamefont {Maunz}}, \ and\ \bibinfo {author} {\bibfnamefont
  {J.}~\bibnamefont {Kim}},\ }\href {\doibase 10.1364/OL.38.004735} {\bibfield
  {journal} {\bibinfo  {journal} {Opt. Lett.}\ }\textbf {\bibinfo {volume}
  {38}},\ \bibinfo {pages} {4735} (\bibinfo {year} {2013})}\BibitemShut
  {NoStop}%
\bibitem [{\citenamefont {Harty}\ \emph {et~al.}(2014)\citenamefont {Harty},
  \citenamefont {Allcock}, \citenamefont {Ballance}, \citenamefont {Guidoni},
  \citenamefont {Janacek}, \citenamefont {Linke}, \citenamefont {Stacey},\ and\
  \citenamefont {Lucas}}]{Harty:2014:rr}%
  \BibitemOpen
  \bibfield  {author} {\bibinfo {author} {\bibfnamefont {T.~P.}\ \bibnamefont
  {Harty}}, \bibinfo {author} {\bibfnamefont {D.~T.~C.}\ \bibnamefont
  {Allcock}}, \bibinfo {author} {\bibfnamefont {C.~J.}\ \bibnamefont
  {Ballance}}, \bibinfo {author} {\bibfnamefont {L.}~\bibnamefont {Guidoni}},
  \bibinfo {author} {\bibfnamefont {H.~A.}\ \bibnamefont {Janacek}}, \bibinfo
  {author} {\bibfnamefont {N.~M.}\ \bibnamefont {Linke}}, \bibinfo {author}
  {\bibfnamefont {D.~N.}\ \bibnamefont {Stacey}}, \ and\ \bibinfo {author}
  {\bibfnamefont {D.~M.}\ \bibnamefont {Lucas}},\ }\href {\doibase
  10.1103/PhysRevLett.113.220501} {\bibfield  {journal} {\bibinfo  {journal}
  {Phys. Rev. Lett.}\ }\textbf {\bibinfo {volume} {113}},\ \bibinfo {pages}
  {220501} (\bibinfo {year} {2014})}\BibitemShut {NoStop}%
\bibitem [{\citenamefont {Wu}\ and\ \citenamefont
  {M{\o}lmer}(2009)}]{Wu:2009:zz}%
  \BibitemOpen
  \bibfield  {author} {\bibinfo {author} {\bibfnamefont {S.}~\bibnamefont
  {Wu}}\ and\ \bibinfo {author} {\bibfnamefont {K.}~\bibnamefont {M{\o}lmer}},\
  }\href {\doibase 10.1016/j.physleta.2009.10.026} {\bibfield  {journal}
  {\bibinfo  {journal} {Phys. Lett. A}\ }\textbf {\bibinfo {volume} {374}},\
  \bibinfo {pages} {34} (\bibinfo {year} {2009})}\BibitemShut {NoStop}%
\bibitem [{\citenamefont {Nielsen}\ and\ \citenamefont
  {Chuang}(2000)}]{Nielsen:2000:tt}%
  \BibitemOpen
  \bibfield  {author} {\bibinfo {author} {\bibfnamefont {M.~A.}\ \bibnamefont
  {Nielsen}}\ and\ \bibinfo {author} {\bibfnamefont {I.~L.}\ \bibnamefont
  {Chuang}},\ }\href@noop {} {\emph {\bibinfo {title} {Quantum Computation and
  Quantum Information}}}\ (\bibinfo  {publisher} {Cambridge University Press},\
  \bibinfo {address} {Cambridge},\ \bibinfo {year} {2000})\BibitemShut
  {NoStop}%
\bibitem [{\citenamefont {Schlosshauer}\ and\ \citenamefont
  {Claringbold}(2014)}]{Schlosshauer:2014:tp}%
  \BibitemOpen
  \bibfield  {author} {\bibinfo {author} {\bibfnamefont {M.}~\bibnamefont
  {Schlosshauer}}\ and\ \bibinfo {author} {\bibfnamefont {T.~V.~B.}\
  \bibnamefont {Claringbold}},\ }in\ Ref.~\cite{Gao:2014:cu},
  pp.\ \bibinfo {pages} {180--194}\BibitemShut {NoStop}%
\bibitem [{\citenamefont {Johansson}\ \emph {et~al.}(2013)\citenamefont
  {Johansson}, \citenamefont {Nation},\ and\ \citenamefont
  {Nori}}]{Johansson:2013:oo}%
  \BibitemOpen
  \bibfield  {author} {\bibinfo {author} {\bibfnamefont {R.}~\bibnamefont
  {Johansson}}, \bibinfo {author} {\bibfnamefont {P.~D.}\ \bibnamefont
  {Nation}}, \ and\ \bibinfo {author} {\bibfnamefont {F.}~\bibnamefont
  {Nori}},\ }\href {\doibase 10.1016/j.cpc.2012.11.019} 
{\bibfield
  {journal} {\bibinfo  {journal} {Comp. Phys. Comm.}\ }\textbf {\bibinfo
  {volume} {184}},\ \bibinfo {pages} {1234} (\bibinfo {year}
  {2013})}\BibitemShut {NoStop}%
\bibitem [{\citenamefont {Breuer}\ and\ \citenamefont
  {Petruccione}(2002)}]{Breuer:2002:oq}%
  \BibitemOpen
  \bibfield  {author} {\bibinfo {author} {\bibfnamefont {H.-P.}\ \bibnamefont
  {Breuer}}\ and\ \bibinfo {author} {\bibfnamefont {F.}~\bibnamefont
  {Petruccione}},\ }\href@noop {} {\emph {\bibinfo {title} {The Theory of Open
  Quantum Systems}}}\ (\bibinfo  {publisher} {Oxford University Press},\
  \bibinfo {address} {Oxford},\ \bibinfo {year} {2002})\BibitemShut {NoStop}%
\bibitem [{\citenamefont {Schlosshauer}(2019)}]{Schlosshauer:2019:qd}%
  \BibitemOpen
  \bibfield  {author} {\bibinfo {author} {\bibfnamefont {M.}~\bibnamefont
  {Schlosshauer}},\ }\href {\doibase 10.1016/j.physrep.2019.10.001} {\bibfield
  {journal} {\bibinfo  {journal} {Phys. Rep.}\ }\textbf {\bibinfo {volume}
  {831}},\ \bibinfo {pages} {1} (\bibinfo {year} {2019})}\BibitemShut {NoStop}%
\bibitem [{\citenamefont {S\o{}rensen}\ and\ \citenamefont
  {M\o{}lmer}(1999)}]{Sorensen:1999:za}%
  \BibitemOpen
  \bibfield  {author} {\bibinfo {author} {\bibfnamefont {A.}~\bibnamefont
  {S\o{}rensen}}\ and\ \bibinfo {author} {\bibfnamefont {K.}~\bibnamefont
  {M\o{}lmer}},\ }\href {\doibase 10.1103/PhysRevLett.82.1971} {\bibfield
  {journal} {\bibinfo  {journal} {Phys. Rev. Lett.}\ }\textbf {\bibinfo
  {volume} {82}},\ \bibinfo {pages} {1971} (\bibinfo {year}
  {1999})}\BibitemShut {NoStop}%
\bibitem [{\citenamefont {Lee}\ \emph {et~al.}(2005)\citenamefont {Lee},
  \citenamefont {Brickman}, \citenamefont {Deslauriers}, \citenamefont
  {Haljan}, \citenamefont {Duan},\ and\ \citenamefont {Monroe}}]{Lee:2005:kk}%
  \BibitemOpen
  \bibfield  {author} {\bibinfo {author} {\bibfnamefont {P.~J.}\ \bibnamefont
  {Lee}}, \bibinfo {author} {\bibfnamefont {K.-A.}\ \bibnamefont {Brickman}},
  \bibinfo {author} {\bibfnamefont {L.}~\bibnamefont {Deslauriers}}, \bibinfo
  {author} {\bibfnamefont {P.~C.}\ \bibnamefont {Haljan}}, \bibinfo {author}
  {\bibfnamefont {L.-M.}\ \bibnamefont {Duan}}, \ and\ \bibinfo {author}
  {\bibfnamefont {C.}~\bibnamefont {Monroe}},\ }\href {\doibase
  10.1088/1464-4266/7/10/025} {\bibfield  {journal} {\bibinfo  {journal} {J.
  Opt. B: Quantum Semiclass. Opt.}\ }\textbf {\bibinfo {volume} {7}},\ \bibinfo
  {pages} {S371} (\bibinfo {year} {2005})}\BibitemShut {NoStop}%
\bibitem [{\citenamefont {Blatt}\ and\ \citenamefont
  {Wineland}(2008)}]{Blatt:2008:uu}%
  \BibitemOpen
  \bibfield  {author} {\bibinfo {author} {\bibfnamefont {R.}~\bibnamefont
  {Blatt}}\ and\ \bibinfo {author} {\bibfnamefont {D.}~\bibnamefont
  {Wineland}},\ }\href@noop {} {\bibfield  {journal} {\bibinfo  {journal}
  {Nature}\ }\textbf {\bibinfo {volume} {453}},\ \bibinfo {pages} {1008}
  (\bibinfo {year} {2008})}\BibitemShut {NoStop}%
\bibitem [{\citenamefont {{Brickman Soderberg}}\ and\ \citenamefont
  {Monroe}(2010)}]{Soderberg:2010:za}%
  \BibitemOpen
  \bibfield  {author} {\bibinfo {author} {\bibfnamefont {K.}~\bibnamefont
  {{Brickman Soderberg}}}\ and\ \bibinfo {author} {\bibfnamefont
  {C.}~\bibnamefont {Monroe}},\ }\href {\doibase 10.1088/0034-4885/73/3/036401}
  {\bibfield  {journal} {\bibinfo  {journal} {Rep. Prog. Phys.}\ }\textbf
  {\bibinfo {volume} {73}},\ \bibinfo {pages} {036401} (\bibinfo {year}
  {2010})}\BibitemShut {NoStop}%
\bibitem [{\citenamefont {Haljan}\ \emph {et~al.}(2005)\citenamefont {Haljan},
  \citenamefont {Brickman}, \citenamefont {Deslauriers}, \citenamefont {Lee},\
  and\ \citenamefont {Monroe}}]{Haljan:2005:km}%
  \BibitemOpen
  \bibfield  {author} {\bibinfo {author} {\bibfnamefont {P.~C.}\ \bibnamefont
  {Haljan}}, \bibinfo {author} {\bibfnamefont {K.-A.}\ \bibnamefont
  {Brickman}}, \bibinfo {author} {\bibfnamefont {L.}~\bibnamefont
  {Deslauriers}}, \bibinfo {author} {\bibfnamefont {P.~J.}\ \bibnamefont
  {Lee}}, \ and\ \bibinfo {author} {\bibfnamefont {C.}~\bibnamefont {Monroe}},\
  }\href {\doibase 10.1103/PhysRevLett.94.153602} {\bibfield  {journal}
  {\bibinfo  {journal} {Phys. Rev. Lett.}\ }\textbf {\bibinfo {volume} {94}},\
  \bibinfo {pages} {153602} (\bibinfo {year} {2005})}\BibitemShut {NoStop}%
\bibitem [{\citenamefont {H\"affner}\ \emph {et~al.}(2003)\citenamefont
  {H\"affner}, \citenamefont {Gulde}, \citenamefont {Riebe}, \citenamefont
  {Lancaster}, \citenamefont {Becher}, \citenamefont {Eschner}, \citenamefont
  {Schmidt-Kaler},\ and\ \citenamefont {Blatt}}]{Haffner:2003:uu}%
  \BibitemOpen
  \bibfield  {author} {\bibinfo {author} {\bibfnamefont {H.}~\bibnamefont
  {H\"affner}}, \bibinfo {author} {\bibfnamefont {S.}~\bibnamefont {Gulde}},
  \bibinfo {author} {\bibfnamefont {M.}~\bibnamefont {Riebe}}, \bibinfo
  {author} {\bibfnamefont {G.}~\bibnamefont {Lancaster}}, \bibinfo {author}
  {\bibfnamefont {C.}~\bibnamefont {Becher}}, \bibinfo {author} {\bibfnamefont
  {J.}~\bibnamefont {Eschner}}, \bibinfo {author} {\bibfnamefont
  {F.}~\bibnamefont {Schmidt-Kaler}}, \ and\ \bibinfo {author} {\bibfnamefont
  {R.}~\bibnamefont {Blatt}},\ }\href {\doibase 10.1103/PhysRevLett.90.143602}
  {\bibfield  {journal} {\bibinfo  {journal} {Phys. Rev. Lett.}\ }\textbf
  {\bibinfo {volume} {90}},\ \bibinfo {pages} {143602} (\bibinfo {year}
  {2003})}\BibitemShut {NoStop}%
\bibitem [{\citenamefont {Leibfried}\ \emph {et~al.}(2003)\citenamefont
  {Leibfried}, \citenamefont {DeMarco}, \citenamefont {Meyer}, \citenamefont
  {Lucas}, \citenamefont {Barrett}, \citenamefont {Britton}, \citenamefont
  {Itano}, \citenamefont {Jelenkovi{\'c}}, \citenamefont {Langer},
  \citenamefont {Rosenband},\ and\ \citenamefont
  {Wineland}}]{Leibfried:2003:mm}%
  \BibitemOpen
  \bibfield  {author} {\bibinfo {author} {\bibfnamefont {D.}~\bibnamefont
  {Leibfried}}, \bibinfo {author} {\bibfnamefont {B.}~\bibnamefont {DeMarco}},
  \bibinfo {author} {\bibfnamefont {V.}~\bibnamefont {Meyer}}, \bibinfo
  {author} {\bibfnamefont {D.}~\bibnamefont {Lucas}}, \bibinfo {author}
  {\bibfnamefont {M.}~\bibnamefont {Barrett}}, \bibinfo {author} {\bibfnamefont
  {J.}~\bibnamefont {Britton}}, \bibinfo {author} {\bibfnamefont {W.~M.}\
  \bibnamefont {Itano}}, \bibinfo {author} {\bibfnamefont {B.}~\bibnamefont
  {Jelenkovi{\'c}}}, \bibinfo {author} {\bibfnamefont {C.}~\bibnamefont
  {Langer}}, \bibinfo {author} {\bibfnamefont {T.}~\bibnamefont {Rosenband}}, \
  and\ \bibinfo {author} {\bibfnamefont {D.~J.}\ \bibnamefont {Wineland}},\
  }\href@noop {} {\bibfield  {journal} {\bibinfo  {journal} {Nature}\ }\textbf
  {\bibinfo {volume} {422}},\ \bibinfo {pages} {412} (\bibinfo {year}
  {2003})}\BibitemShut {NoStop}%
\bibitem [{\citenamefont {Lee}\ \emph {et~al.}(2016)\citenamefont {Lee},
  \citenamefont {Smith}, \citenamefont {Richerme}, \citenamefont {Neyenhuis},
  \citenamefont {Hess}, \citenamefont {Zhang},\ and\ \citenamefont
  {Monroe}}]{Lee:2016:in}%
  \BibitemOpen
  \bibfield  {author} {\bibinfo {author} {\bibfnamefont {A.~C.}\ \bibnamefont
  {Lee}}, \bibinfo {author} {\bibfnamefont {J.}~\bibnamefont {Smith}}, \bibinfo
  {author} {\bibfnamefont {P.}~\bibnamefont {Richerme}}, \bibinfo {author}
  {\bibfnamefont {B.}~\bibnamefont {Neyenhuis}}, \bibinfo {author}
  {\bibfnamefont {P.~W.}\ \bibnamefont {Hess}}, \bibinfo {author}
  {\bibfnamefont {J.}~\bibnamefont {Zhang}}, \ and\ \bibinfo {author}
  {\bibfnamefont {C.}~\bibnamefont {Monroe}},\ }\href {\doibase
  10.1103/PhysRevA.94.042308} {\bibfield  {journal} {\bibinfo  {journal} {Phys.
  Rev. A}\ }\textbf {\bibinfo {volume} {94}},\ \bibinfo {pages} {042308}
  (\bibinfo {year} {2016})}\BibitemShut {NoStop}%
\bibitem [{\citenamefont {Brydges}\ \emph {et~al.}(2019)\citenamefont
  {Brydges}, \citenamefont {Elben}, \citenamefont {Jurcevic}, \citenamefont
  {Vermersch}, \citenamefont {Maier}, \citenamefont {Lanyon}, \citenamefont
  {Zoller}, \citenamefont {Blatt},\ and\ \citenamefont
  {Roos}}]{Brydges:2019:aa}%
  \BibitemOpen
  \bibfield  {author} {\bibinfo {author} {\bibfnamefont {T.}~\bibnamefont
  {Brydges}}, \bibinfo {author} {\bibfnamefont {A.}~\bibnamefont {Elben}},
  \bibinfo {author} {\bibfnamefont {P.}~\bibnamefont {Jurcevic}}, \bibinfo
  {author} {\bibfnamefont {B.}~\bibnamefont {Vermersch}}, \bibinfo {author}
  {\bibfnamefont {C.}~\bibnamefont {Maier}}, \bibinfo {author} {\bibfnamefont
  {B.~P.}\ \bibnamefont {Lanyon}}, \bibinfo {author} {\bibfnamefont
  {P.}~\bibnamefont {Zoller}}, \bibinfo {author} {\bibfnamefont
  {R.}~\bibnamefont {Blatt}}, \ and\ \bibinfo {author} {\bibfnamefont {C.~F.}\
  \bibnamefont {Roos}},\ }\href {\doibase 10.1126/science.aau4963} {\bibfield
  {journal} {\bibinfo  {journal} {Science}\ }\textbf {\bibinfo {volume}
  {364}},\ \bibinfo {pages} {260} (\bibinfo {year} {2019})}\BibitemShut
  {NoStop}%
\bibitem [{\citenamefont {Maier}\ \emph {et~al.}(2019)\citenamefont {Maier},
  \citenamefont {Brydges}, \citenamefont {Jurcevic}, \citenamefont {Trautmann},
  \citenamefont {Hempel}, \citenamefont {Lanyon}, \citenamefont {Hauke},
  \citenamefont {Blatt},\ and\ \citenamefont {Roos}}]{Maier:2019:uu}%
  \BibitemOpen
  \bibfield  {author} {\bibinfo {author} {\bibfnamefont {C.}~\bibnamefont
  {Maier}}, \bibinfo {author} {\bibfnamefont {T.}~\bibnamefont {Brydges}},
  \bibinfo {author} {\bibfnamefont {P.}~\bibnamefont {Jurcevic}}, \bibinfo
  {author} {\bibfnamefont {N.}~\bibnamefont {Trautmann}}, \bibinfo {author}
  {\bibfnamefont {C.}~\bibnamefont {Hempel}}, \bibinfo {author} {\bibfnamefont
  {B.~P.}\ \bibnamefont {Lanyon}}, \bibinfo {author} {\bibfnamefont
  {P.}~\bibnamefont {Hauke}}, \bibinfo {author} {\bibfnamefont
  {R.}~\bibnamefont {Blatt}}, \ and\ \bibinfo {author} {\bibfnamefont {C.~F.}\
  \bibnamefont {Roos}},\ }\href {\doibase 10.1103/PhysRevLett.122.050501}
  {\bibfield  {journal} {\bibinfo  {journal} {Phys. Rev. Lett.}\ }\textbf
  {\bibinfo {volume} {122}},\ \bibinfo {pages} {050501} (\bibinfo {year}
  {2019})}\BibitemShut {NoStop}%
\bibitem [{Note2()}]{Note2}%
  \BibitemOpen
  \bibinfo {note} {See Ref.~\cite {Dass:1999:az} for an analysis of wave-packet
  spreading during a protective measurement.}\BibitemShut {Stop}%
\end{thebibliography}

%

\end{document}